\documentclass[journal,twoside,web]{ieeecolor}

\usepackage{etoolbox}
\makeatletter
\@ifundefined{color@begingroup}%
{\let\color@begingroup\relax
\let\color@endgroup\relax}{}%
\def\fix@ieeecolor@hbox#1{%
\hbox{\color@begingroup#1\color@endgroup}}
\patchcmd\@makecaption{\hbox}{\fix@ieeecolor@hbox}{}{\FAILED}
\patchcmd\@makecaption{\hbox}{\fix@ieeecolor@hbox}{}{\FAILED}

\usepackage{pifont}
\usepackage{tmi}
\usepackage{cite}
\usepackage{amsmath,amssymb,amsfonts}
\usepackage{algorithmic}
\usepackage{graphicx}
\usepackage{textcomp}
\usepackage{epsfig}
\usepackage{booktabs}
\usepackage[normalem]{ulem}
\usepackage{multirow}
\usepackage[ruled,linesnumbered]{algorithm2e}
\usepackage{comment}
\usepackage{overpic}
\usepackage{bm}

\newcommand\HUI[1]{ }
\newcommand\Special[1]{ }
\newcommand\ourmodel{LInFBP}

\def\BibTeX{{\rm B\kern-.05em{\sc i\kern-.025em b}\kern-.08em
    T\kern-.1667em\lower.7ex\hbox{E}\kern-.125emX}}
\begin{document}
\title{Continuous Filtered Backprojection by Learnable Interpolation Network}
\author{\makebox[\linewidth]{\parbox{\dimexpr\textwidth+1cm\relax}{\centering Hui Lin, Dong Zeng, Qi Xie, Zerui Mao, Jianhua Ma, and Deyu Meng}}
\thanks{H.~Lin,~Q.~Xie,~and~D.~Meng are with the School of Mathematics and Statistics, Xi'an Jiaotong University, Shaanxi, 710049, China. }
\thanks{D.~Zeng,~and~Z.~Mao are with the School of Biomedical Engineering, Southern Medical University, Guangzhou 510515, China.}
\thanks{J.~Ma is with the School of Life Science and Technology, Xi'an Jiaotong University, Shaanxi, 710049, China.}
}

\maketitle

\begin{abstract}
Accurate reconstruction of computed tomography (CT) images is crucial in medical imaging field. 
However, there are unavoidable interpolation errors in the backprojection step of the conventional reconstruction methods, i.e., filtered-back-projection based methods, which are detrimental to the accurate reconstruction.
In this study, to address this issue, we propose a {novel deep learning model, named Leanable-Interpolation-based FBP or \ourmodel{} shortly,} to enhance the reconstructed CT image quality, which achieves learnable interpolation in the backprojection step of filtered backprojection (FBP) {and alleviates the interpolation errors.} 
Specifically, in the proposed \ourmodel{}, we formulate every local piece of the  latent continuous function of discrete sinogram data  as a linear combination of selected basis functions, and learn this  continuous function by  exploiting a deep network to predict the linear combination coefficients. 
Then, the learned  latent continuous function is exploited for interpolation in backprojection step, which first time takes the advantage of deep learning for the interpolation in FBP. 
Extensive experiments, which encompass diverse CT scenarios, demonstrate the effectiveness of the proposed \ourmodel{} in terms of enhanced reconstructed image quality, plug-and-play ability and generalization capability.
\end{abstract}


\section{Introduction}
\label{sec:introduction}
Radon transform has been implemented in various applications due to its efficiency and effectiveness, especially in computed tomography (CT) field~\cite{cormack1963representation, hsieh2003computed}. The inverse Radon transform, a continuous integral operation, effectively reconstructs a continuous image distribution from a corresponding sinogram distribution. However, it's important to note that in practical applications, CT images are typically discretized for numerical processing. In such cases, a discrete approximation of the inverse Radon transform algorithm, \textit{i.e.}, filtered backprojection (FBP) algorithm~\cite{shepp1974fourier}, is often employed for CT reconstruction in real-world scenarios. Specifically, FBP correlates discrete image domain with sinogram data and has become the predominant analytic reconstruction method due to its computational efficiency and numerical stability. However, due to the finite detector bins, insufficient projection views or unavoidable noise, the quality of FBP reconstructed images tends to degrade remarkably without adequate treatments.
%

Many improvement schemes for original FBP have been proposed to enhance the performance, in the past few years. Typical instances encompass sinogram pre-processing methods~\cite{xie2017robust, wang2014pet, lee2018deep, li2019promising, dong2019sinogram, tan2019sharpness}, 
filter learning methods~\cite{he2020radon, zeng2013filtered, zeng2014noise}, 
sophisticated interpolation methods~\cite{la1998spline, horbelt2002discretization, entezari2012box}, and
view-by-view backprojection domain processing methods~\cite{tao2019revisit, tao2019vvbp, chen2017low, li2020sacnn, han2022dual}.

Among them, developing learnable backprojection strategies has attracted a growing interest. For instance, Zhu et al.~\cite{zhu2018image} developed an automated transform by manifold approximation (AUTOMAP) to achieve an end-to-end reconstruction for Radon inversion. Later on, He et al.~\cite{he2020radon} proposed an inverse Radon transform approximation (iRadonMAP) network by constructing a learned sinusoidal backprojection operation for CT image reconstruction to decrease computational burden in the AUTOMAP. More recently, Tao et al.~\cite{tao2021learning} investigated to process critical intermediate components in the FBP implementation to reconstruct high-quality CT images at various dose levels.

Although these methods have achieved a certain degree of success in constructing learnable FBP, there is still room for performance improvements. The first aspect is to improve parameter efficiency and generalization capability within the existing learnable FBP framework. For example, the backprojection mapping in iRadonMap~\cite{he2020radon} is indeed a partially connected linear layer, which simply assigns learnable weights to the transformations between all sinogram pixels and image pixels. This approach results in a large number of learnable parameters ($M \times 512 \times 512$, where $M$ is the number of sampling views), leading to significant parameter costs and computational demands that could potentially limit the final reconstruction performance. 
This substantial parameter costs could also hinder the model's generalization capability, 
which certainly constrains the plug-and-play functionality as well.


Based on the aforementioned analysis, we can conclude that the key to construct an effective learnable backprojection module is to reduce the unnecessary learnable settings, \textit{i.e.}, making optimal utilization of the accurate components in original FBP to the possibly great extent and only enhance the inaccurate parts through deep learning approaches. 

\begin{figure}[t]
  \begin{center}
  \includegraphics[width = 1\linewidth]{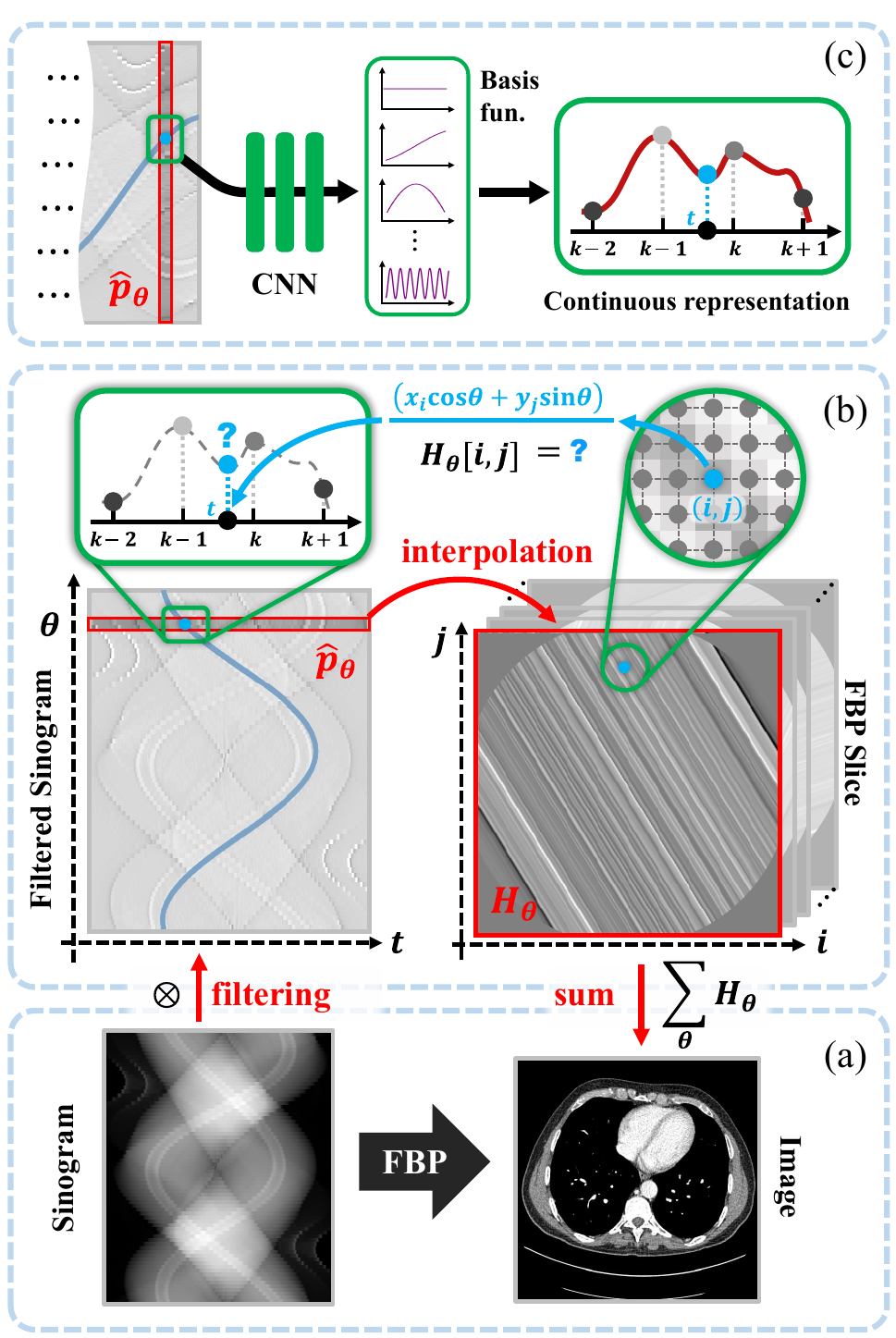}
   \caption{An overview of the FBP process (taking the parallel-beam imaging geometry as instance). The interpolation issue arises when sampling the value of a FBP slice from the filtered sinogram.}
   \label{fig:intro}
   \end{center}\vspace{-3mm}
\end{figure}

Actually, the deduction of inverse Radon transform is entirely exact in the continuous domain. Errors \HUI{occur} in FBP only due to the discretization implementation. Specifically, the FBP algorithm typically consists of three steps: filtering, view-by-view backprojection, and summation, as shown in Fig. \ref{fig:intro}. In the filtering step, the FBP conducts discrete filtration on the sinogram data to remove blur effects. And then in the view-by-view backprojection step, each view of the filtered sinogram is backprojected into the image domain. Finally in the summation step, all the sinogram values picked up at each view are summarized along $\theta$ direction to obtain the reconstructed CT image. Specifically, the approximation errors  primarily exist in two components: (1) the selection of discrete filter and (2) the interpolation in process view-by-view backprojection. \HUI{For example, as shown in Fig.~\ref{fig:intro}(b), it is difficult to determine an appropriate value for each pixel of the view-by-view backprojection tensor because the image and sinogram are in different coordinate systems and the coordinates corresponding to the projected pixels are generally non-integer values. The values of these coordinates are thus not directly detected by the detector bin and can only be achieved by numerical interpolation. Previous research has provided clear justifications that optimizing the interpolation process during the backprojection step can substantially improve reconstruction accuracy and effectively reduce reconstruction errors \cite{tang2018effect}.} Although current learnable FBP methods~\cite{he2020radon, he2021downsampled} take advantage of using specific deep-network layers to represent filtering steps, they rarely consider interpolation steps in the network construction, which limits their further performance improvements. 

In this paper, to alleviate the discretization error and improve the reconstruction performance in current learnable FBP methods, we focus on designing a continuous representation network for learning the continuous sinogram with a small number of parameters, as shown in Fig. \ref{fig:intro}(c). This continuous representation network can be treated as a learnable interpolation in the view-by-view backprojection process, and is expected to further mitigate the approximation error in the interpolation process of FBP as compared with traditional interpolation. To the best of our knowledge, this should be the first study that adds effort to characterize the discrete sinogram by a learnable continuous representation network. Specifically, this paper mainly makes the following four contributions.

\begin{enumerate}
\item We present a novel continuous representation network for FBP, named \ourmodel{} (the abbreviation of Learnable-Interpolation-based FBP). The proposed \ourmodel{} first time customizes to the task of detector bin interpolation with learnable local continuous representations (LCR), {which mitigates the unavoidable interpolation error}.   
\item By replacing the imprecise conventional interpolation operation within the FBP implementation while striving to maintain the other accurate operations, the proposed \ourmodel{} largely inherits the computational flow and physical interpretability of FBP, thus demonstrating less parameters and superior generalization performance, compared to other deep-learning-based methods. Particularly, the proposed method surpasses traditional FBP, even when there are significant changes in the imaging geometry of the testing data. {In contrast, most of other deep-learning-based methods would fail to achieve satisfactory results in such case.} 
\item The proposed \ourmodel{} can be seamlessly integrated with the majority of existing FBP-based CT reconstruction methods in a plug-and-play manner. In this work, we study the feasibility of \ourmodel{} by integrating it into two state-of-the-art (SOTA) deep learning-based CT reconstruction models. By replacing the back projection module, the reconstruction results showcase the adaptation and generalization capabilities of \ourmodel{}.
\item We have validated the effectiveness of propose model in various scenarios, including noise-free, low dose, sparse view, and noisy data. Compared to the traditional nearest neighbor and linear interpolation, {the proposed method shows improvements in all evaluation metrics}. Besides, adequate experiments show that integrating the proposed module into SOTA methods can obviously help improving the performance.
\end{enumerate}

The rest of this paper is organized as follows. Section~\ref{sec:methodology} elaborates the proposed method. Section~\ref{sec:setup} and~\ref{sec:results} presents the experimental details and the results. And Section~\ref{sec:conclusion} draws the conclusions.

\section{Methodology}\label{sec:methodology}

The key idea of the proposed methodology is to design a learnable view-by-view interpolation regime. Therefore, we start by providing a comprehensive overview of the theoretical foundations of the Radon transform, emphasizing the interpolation errors that occur during the backprojection process. Then, we introduce our proposed approach, a local continuous representation for filtered sinogram, which naturally lead to better interpolation for the view-by-view backprojection.

We take the parallel-beam imaging geometry as instance to introduce our method, and it is worth noting that the cases of fan-beam and cone-beam imaging can be deduced similarly.

\subsubsection{Inverse Radon transform in continuous domain}
Mathematically, the forward projection and backprojection based on the Radon transform can be formulated as follows~\cite{tao2021learning}:
\begin{equation}\label{eq:fp}
    p_\theta(t) = \iint_{\mathbb{R}^2}  \mu(x,y)\delta(x \cos \theta + y \sin \theta - t) dxdy,
\end{equation}
where $p_\theta(t)$ denotes the projection sinogram with the detector bin at the position $t$ and the exposure angle $\theta$. $\mu(x,y)$ denotes the image with 2D coordinate $(x,y)$. The goal for CT image reconstruction is to estimate $\mu$ with observed $p$. By exploiting  Fourier slice theorem, the  exact solution for the problem can be theoretically deduced  in the following inverse Radon transform form \cite{gengsheng2010medical}:
\begin{equation}\label{eq:bp}
    \mu(x,y) = \int^{2\pi}_0 \! \int^\infty_{-\infty}\! |\omega| P_\theta(\omega) e^{-2\pi i  \omega (x \cos \theta + y \sin \theta)} d\omega d\theta,
\end{equation}
where $P_\theta(\omega)$ is obtained by applying 1D Fourier transform on $p_\theta(t)$.
In general, (\ref{eq:bp}) reconstructs the sinogram to the image, where each point in the reconstruction integrates the projection along a specific sinusoidal curve, as shown in Fig.~\ref{fig:intro}(b).

\subsubsection{Discrete approximation in practice}

In practice, due to the limited number of  detectors and views, one can  only perform the discrete approximation of (\ref{eq:bp}), \textit{i.e.}, the FBP algorithm.

Specifically, the calculation of (\ref{eq:bp}) is usually decomposed into the following 3 stages before discretization:
\begin{equation}\label{3stage}
  \left\{ \begin{array}{cll}
            \tilde{p}_\theta(t) &=& \int^\infty_{-\infty} |\omega| P_\theta(\omega) e^{-2\pi i  \omega t} d\omega,  \\
             \tilde{H}_\theta(x,y) &=& \tilde{p}_\theta(x \cos \theta + y \sin \theta),\\
           \mu(x,y) &=& \int^{2\pi}_0 \tilde{H}_\theta(x,y) d\theta.
          \end{array}  \right.
\end{equation}
The discretization of these 3 stages are correlated to the 3 key steps in FBP algorithm, respectively, \textit{i.e.}, filtering, view-by-view backprojection, and summing. One can view Fig. \ref{fig:intro} for an overall understanding of these 3 steps.

In the filtering step, the calculation of  $\tilde{p}_\theta(t)$ can be rationally approximated by a filtering operation \cite{wurfl2016deep, syben2017precision},
\begin{equation}\label{eq:filter}
    \hat{\bm{p}}_\theta =\bm{q}\otimes \bm{p_\theta},
\end{equation}
where $\otimes$ denotes the convolution operation, $\bm{p}_\theta\in \mathbb{R}^N$ represents the detected sinogram at $\theta$ exposure angle, $\hat{\bm{p}}_\theta\in \mathbb{R}^N$ denotes the corresponding filtered sinogram, and $\bm{q}$ is the specifically selected filter.

In the view-by-view backprojection step, the FBP slice $\bm{H}_{\theta}\in \mathbb{R}^{h\times w}$ at each exposure angle must be estimated from  $\hat{\bm{p}}_\theta$,  where  $h$ and $w$ are the height and width of the to-be-reconstructed  CT image, respectively.
Since for a pixel at position $(x,y)$, the index $x \cos \theta + y \sin \theta$ is always not an integer, interpolation on the discrete  $\hat{\bm{p}}_\theta$ will be necessary and $\bm{H}_{\theta}$ is calculated as following:
\begin{equation}\label{eq:H_theta}
    \bm{H}_\theta[{i,j}] = \tilde{\varphi}_\theta(x_i \cos \theta + y_j \sin \theta),
\end{equation}
where $\tilde{\varphi}_\theta(t)$ is the interpolation function of $\hat{\bm{p}}_\theta$,  and $(x_i, y_j)$ is the coordinate of the $j$-th column and $i$-th row in $\bm{H}_\theta$.

In the summing step, the calculation of the integral along the continuous exposure angle domain can be naively approximated by summing along limited exposure angles, \textit{i.e.},
\begin{equation}\label{eq:sum}
    \bm{I} = \frac{2\pi}{M} \sum_{m=1}^M \bm{H}_{\theta_m},\  \theta_m=\frac{2\pi m}{M},
\end{equation}
where $\bm{I}\in \mathbb{R}^{h\times w}$ denotes the to-be-reconstructed CT image and $M$ is the number of exposure  views.

In practice, approximation errors will occur in all of (\ref{eq:filter}), (\ref{eq:H_theta}) and (\ref{eq:sum}). Therefore, reducing approximation errors in these stages would be the key issues to enhance FBP capability. Since it is rational to approach integral by summming, the room for improvements should lie in (\ref{eq:filter}) and (\ref{eq:H_theta}). In recent years, inspired by the success of deep learning approach, neural network has been introduced for replacing the filter operator in (\ref{eq:filter}), and shown to be helpful for improving the performance of FBP~\cite{he2020radon}.
This is quite intuitive, since filtering operation can be viewed as one layer of CNN and learnt by deep learning technique. In this paper we will focus on a more  challenging task, \textit{i.e.},  reducing the approximation errors in (\ref{eq:H_theta}).


\subsection{Learnable view-by-view backprojection framework}

\begin{figure*}[t]
  \begin{center}
  \includegraphics[width = 1\linewidth]{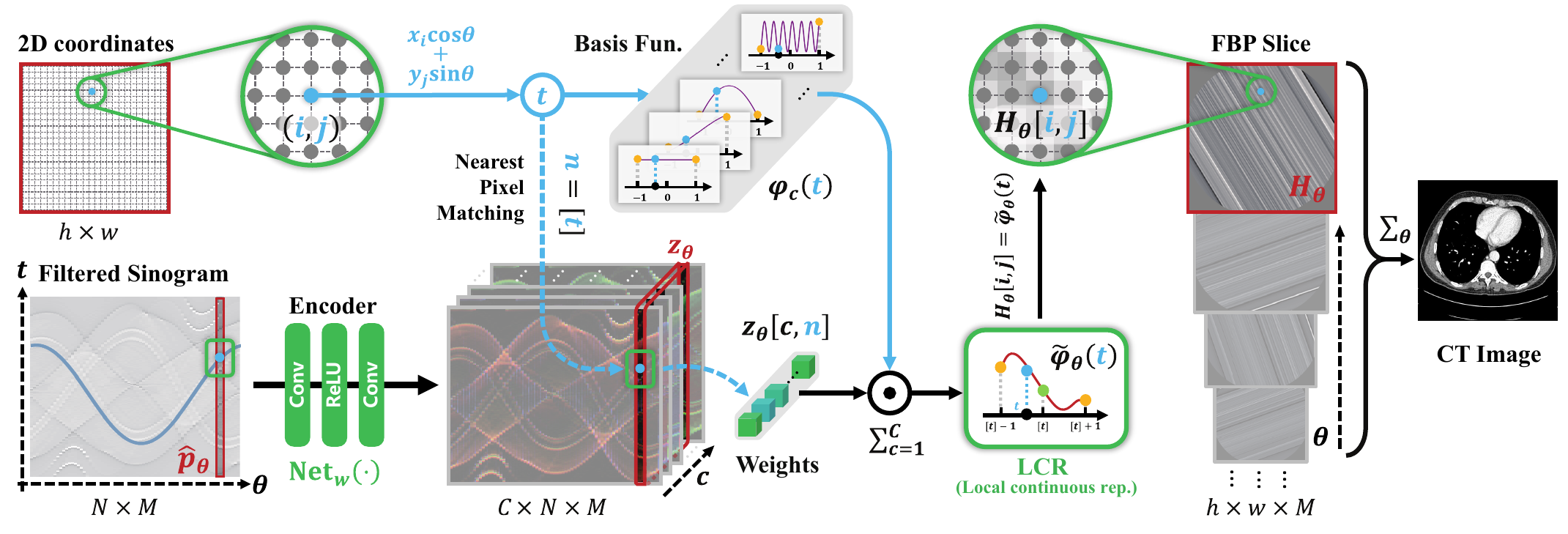}
  \vspace{-2mm}
   \caption{The total framework of our model. Utilizing a trained deep learning model to predict composition weights, we achieve a continuous representation of the sinogram by combining a series of basis functions. When sampling sinogram values from positions not covered by existing detectors, we can directly obtain these transformed values from this continuous representation using relative coordinates as queries. This approach effectively mitigates interpolation errors.}
   \label{fig:model}
   \end{center}\vspace{-2mm}
\end{figure*}

Instead of using fixed interpolation method, such as nearest, linear or cubic interpolation, to calculate $\tilde{\varphi}_\theta$ in (\ref{eq:H_theta}), we propose an adaptive approach that provides a proper approximation to $\tilde{p}_\theta(t)$ in the context of data-driven. Specifically, we construct a learnable continuous representation method along with a convolution network to learn the interpolation function for $\hat{\bm{p}}_\theta$. One can see Fig. \ref{fig:model} for an easy understanding of the proposed framework.

As shown in Fig.~\ref{fig:model},  learnable local continuous representation (LCR) is the key  for predicting filtered sinogram values at arbitrary given coordinate. 
In this study we exploit a light-weight continuous representation framework, which can be formulated as:
\begin{equation}\label{general}
  \tilde{\varphi}(t) = \sum_{c = 1}^C w_c \cdot\varphi_{c}(t),
\end{equation}
where  $\tilde{\varphi}$ is a to-be-learnt function,  
$\varphi_c$, $c = 1,2,\cdots, C$  denote $C$ preset basis functions which will be discussed in details in the later sections, and $w_c$ is the $c$-th representation weight.  It is easy to find that we can learn different $\tilde{\varphi}$ by learning different $w_c$.

Following this framework, 
we propose the LCR for filter sinogram at each view respectively:
\begin{equation}\label{eq:svomega}
    \tilde{\varphi}_\theta(t) = \sum_{c = 1}^C \bm{z}_\theta\left[c, [ t ]\right] \cdot\varphi_c(t-[ t ]),
\end{equation}
where $\tilde{\varphi}_\theta$ denotes the learned interpolation function for the $\theta$ view, with $t$ being its argument; $[t]$ denotes the nearest integer to $t$;
$\bm{z}_\theta\in\mathbb{R}^{C\times N}$ represents an encoded vector that collects local information from the sinogram, which also play the role of weights for the LCR.
By substituting (\ref{eq:svomega}) into (\ref{eq:H_theta}) as  interpolation function, the sinogram value at  $t = x_i \cos \theta + y_j \sin \theta$ can be represented according to the information provided by its neighboring detectors, \textit{i.e.}, $[ t ]$-th latent encoded vector\footnote{When
$t$ moves, the selection of $[t]$ may suddenly switch and cause discontinuous issue on $\tilde{\varphi}_\theta(t)$. In practice, it has been shown that  the local ensemble technique \cite{chen2021learning} can well  address this issue.}. 

For the latent encoded vector $\bm{z}_\theta$, we predict $\bm{z}=[\bm{z}_{\theta_1}, \bm{z}_{\theta_2}, \cdots, \bm{z}_{\theta_M}]\in \mathbb{R}^{C \times N \times M}$  in deep learning manner, \textit{i.e.},
\begin{equation}\label{eq:Znetwork}
  \bm{z} = \operatorname{Net}_w(\hat{\bm{p}}),
\end{equation}
where $\hat{\bm{p}}= \left[ \hat{\bm{p}}_{\theta_1}, \hat{\bm{p}}_{\theta_2}, \cdots, \hat{\bm{p}}_{\theta_M}\right]$ corresponds to the filtered sinogram data matrix, which is with the size of $N\times M$, and  $ \operatorname{Net}_w$ is a learnable network whose parameters denoted as $w$. Combining (\ref{eq:H_theta}), (\ref{eq:sum}), (\ref{eq:svomega}) and (\ref{eq:Znetwork}), the end-to-end CT reconstruction model can be formulated as:
\begin{equation}\label{eq:allModel}\begin{split}
  &I[i,j] = \frac{2\pi}{M}\sum_{c,m}\operatorname{Net}_w(\hat{\bm{p}})[c,n,m]\cdot\varphi_c(t - n),\\
  \end{split}
\end{equation}
where $t = x_i \cos \theta_m + y_j \sin \theta_m$, and $n = [ t ]$.
In the following we call this network as \ourmodel{} (the abbreviation of Learnable-Interpolation-based FBP). Besides, for the sake of convenience, we rewrite (\ref{eq:allModel}) as \begin{equation}\label{simple}
                         \hat{\bm{I}} = \operatorname{LInFBP}_w(\hat{\bm{p}}).
                       \end{equation} 

One can exploit different deep networks as  $\operatorname{Net}_w$. In our approach, we simply set $ \operatorname{Net}_w$ as a stack of two convolutional layers and a ReLU function, in the from of `Conv. + ReLU + Conv.', where the convolution layers are set as 1D convolutions. We empirically find that such a simple network framework can already help greatly improve CT reconstruction result, as comparing with traditional FBP and other previous learnable FBP methods, with unsubstantial  additional computational cost. It should be noted that one can exploit more complex networks to further improve the $\bm{z}$ prediction performance, based on the requirement in practice.

Specifically, when directly using \ourmodel{} for CT image reconstruction (the most fundamental usage of \ourmodel{}), the training loss is:
\begin{equation}\label{eq:loss}
  \min_w \sum_j\nolimits \mathcal{L}\left( \operatorname{LInFBP}_w(\hat{\bm{p}_j}), \bm{I}_j\right),
\end{equation}
where $\bm{I}_j$ is the reference image of each training sample $j$, $\mathcal{L}$ is a chosen loss function, and $w$ is the only parameter needed to be learned. In practice, one can readily integrate \ourmodel{} into current CT reconstruction method by replacing the backprojection module in a plug-and-play manner.

\subsection{Selection of basis function set}
It is evident that selecting basis function set $\{\varphi_c(t)\}_{c=1}^C$ is important for the propose LCR in (\ref{eq:svomega}). 
In this paper, we exploit two different basis function sets for respectively constructing LCR and comparing their performance. The first one is the Fourier basis function set, which has been used in many previous LCR methods \cite{song2023ope,xie2022fourier}. The second one is a new proposed function set, named the linear basis function set, which can greatly reduce the computational cost when performing LCR in CT backprojection.
 
\subsubsection{Fourier basis function set}
It has been shown in previous methods that the linear combination of Fourier basis functions can approach commonly-used functions with very high precision by setting appropriate max frequency for the basis function set \cite{song2023ope,xie2022fourier}. Generally,  as shown in Fig. \ref{fig:baisfun}(a), it's with the following formulation:
\begin{equation}\label{Fourier}
   \left\{\varphi_{c}(t)\!=\!\sin(ct), \varphi_{c+\!1}(t)\! =\! \cos(c t)\big|c \!=\! 0,2,\cdots\!,2k\right\}\!,\!
\end{equation}
where $k$ denotes the preset max frequency for the Fourier basis function set, and the total number of the basis functions is $C = 2k+1$ (the term $\sin(0 \cdot t)$ consistently equals to $0$, which can be ignored in practice).
We then name the learnable FBP method with this Fourier basis based LCR as F-\ourmodel{}.

\subsubsection{Linear basis function set}
In practice, the high resolution of the target reconstructed image still necessitates the sampling of a large number of positions, typically $M \times 512 \times 512$, where $M$ is the number of views. 
This makes LCR (\ref{eq:svomega}) with Fourier function set relatively computational expensive. Particularly, we need to compute the summation of $2k+1$ large tensor ($\varphi_{c}(x_i \cos \theta + y_j \sin \theta)$ for all $i$, $j$ and $\theta$, \textit{i.e.}, a tensor of size $M \times 512 \times 512$) to obtain all the $\bm{H}_\theta$.

\begin{figure}[t]
  \begin{center}\hspace{-2mm}
  \includegraphics[width = 0.97\linewidth]{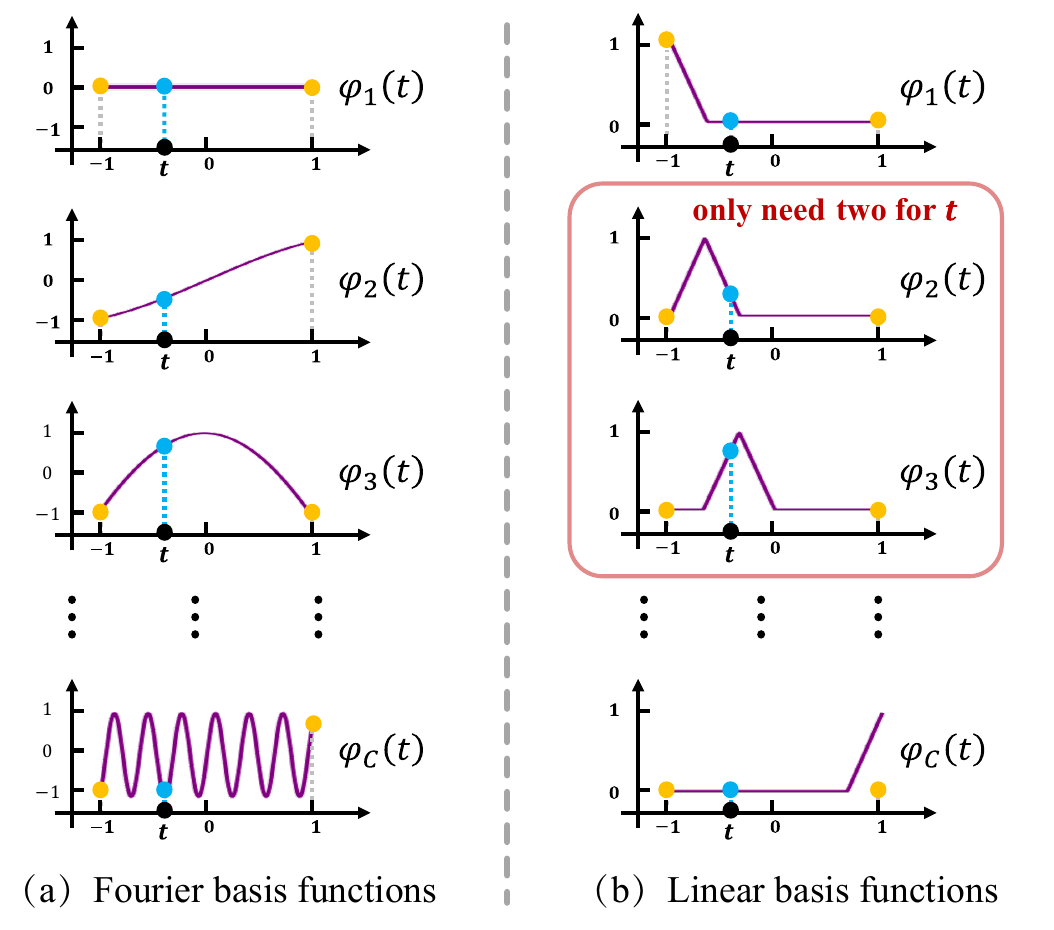}\vspace{-2mm}
   \caption{Illustrations of the Fourier  and  linear basis function set.}
   \label{fig:baisfun}
   \end{center}\vspace{-4mm}
\end{figure}

To address this challenge, we propose a new basis set for lightweight LCR named as the linear kernel basis set. 
To be specific, we propose the following basis function set for LCR (\ref{eq:svomega}). It is in the following form:
\begin{equation}\label{Linear}
\begin{split}
\left\{\varphi_{c}(t) \!=\! \max\!\left(1\!-\! \left|kt \!-\! (c\!-\!k\!-\!1\!)\right|\!, 0\right)\!\big| c\!=\!1,\!\cdots\!,2k\!\!+\!\!1\right\},
\end{split}
\end{equation}
where $k$ is a preset number, which controls the number of functions in this function set ($C = 2k+1$).  One can refer to Fig. \ref{fig:baisfun}(b) for easy understanding this function set. 


It should be noted that for any $t\in[-1,1]$, we actually only need two of the basis functions for the LCR in (\ref{general}). Specifically, denoting $\lceil \cdot \rceil$ and $\lfloor \cdot \rfloor$ as ceiling and floor functions, respectively, for a given $t\in[-1,1]$, there exists an integer $\bar{c}$ \textit{st.} $t\in \left[\frac{\bar{c}-k-1}{k}, \frac{\bar{c}-k}{k}\right]$. It is easy to find ${\bar{c}-k-1} = {\lfloor kt \rfloor} $ and ${\bar{c}-k} = {\lceil kt \rceil} $. Thus we have 
\begin{equation}\label{phic}
  \varphi_c(t) = \left\{ \begin{array}{ccc}
                           {\lceil kt \rceil} - kt & \mbox{if} & c = \bar{c} \\
                           kt - \lfloor kt \rfloor & \mbox{if} & c = \bar{c}+1 \\
                           0 & \mbox{if} & \mbox{otherwise}
                         \end{array} \right..
\end{equation}
Then, by substituting (\ref{phic}) into  (\ref{general}), the LCR become
\begin{equation}\label{eq:linear_more}
\begin{aligned}
\tilde{\varphi}(t) &= \sum_{c = 1}^C w_c \cdot\varphi_{c}(t) = w_{\bar{c}} \cdot\varphi_{\bar{c}}(t) + w_{\bar{c}+1}\cdot\varphi_{\bar{c}+1}(t) \\
&= w_{\bar{c}} \cdot \left(\lceil kt \rceil\!-\!kt\right) + w_{\bar{c}+1} \cdot \left(kt\!-\!\lfloor kt \rfloor\right),
\end{aligned}
\end{equation}
which is in a relatively simple calculation form.
Due to this, we no longer need to compute the summation of $2k+1$ large tensors just as traditional LCR like F-\ourmodel{} did,
but only  need to compute the summation of $2$ tensors. This greatly reduces the computational cost.
We name this linear kernel basis based learnable FBP method as L-\ourmodel{}.

\section{Experimental Setup}\label{sec:setup}

\subsection{Dataset}~\label{sec:aapm}
In this work, to validate and evaluate the performance of the proposed \ourmodel{}, two datasets are used, i.e., the AAPM Mayo Clinic dataset~\cite{mccollough2016tu} and the Clinical LDCT dataset. Both of them are acquired by Siemens CT scanner.

1) AAPM Mayo Clinic dataset: The experiments are conducted on the AAPM Mayo Clinic dataset authorized for the `2016 NIH-AAPM-Mayo Clinic Low Dose CT Grand Challenge'. The dataset contains whole-body CT image from ten patients, which is acquired by Siemens SOMATOM Definition AS+CT scanner with voltage levels of 120 kVp or 100 kVp, and exposure within the range of 200 to 500 mA. In the experiments, the data acquired with 120 kVp and 200 effective mAs were chosen as ground-truth. {Nine patients data (4,791 slices) are used for network training, and the remaining data (560 slices) from one patient are used for test.} To fully demonstrate the reconstruction performance of the proposed methods in the low-dose CT imaging, five different noise levels are simulated from the corresponding ground-truth data based on our previous study, i.e., 25\% of full dose  (quarter-dose), 25\% of 1160 sampling-views (quarter-view), 25\% of 1160 sampling-views (quarter-view) + 25\% of full dose (quarter-dose), 145 sparse views (145 views), 100 sparse views (100 views), and 72 sparse views (72 views). {The fan beam geometry is used in the simulation study with the following geometry parameters: the voxel grid is $512 \times 512$ with voxel size of 0.6641 mm, source-to-detector distance is 1058.6 mm, source-to-isocenter distance is 595.0 mm, number of detector bins is 736, and detector bin width is 1.3696 mm.} Using the ground-truth as the training label, the respective network was trained across each dose level to develop the corresponding network.

2) Clinical LDCT dataset: To further demonstrate the potential of the proposed methods in real scenarios, real LDCT data of 10 individual participants at the local hospital are collected. Informed consent was obtained from all individual participants included in the study. These LDCT images are acquired with a Siemens Somatom Definition AS CT scanner with 120 kVp and 60 mAs. Specifically, 960 real LDCT images were utilized to generate sinograms via forward projection and the geometry parameters are the same as those in the AAPM Mayo Clinic dataset, and then they are fed into the well-trained F-\ourmodel{} and L-\ourmodel{}, respectively, to reconstruct the final CT images.


\subsection{Implementation Details}

Our model is implemented using PyTorch toolbox~\cite{paszke2017automatic} and Astra~\cite{van2015astra} toolbox on the NVIDIA GeForce RTX 3090 GPU. The RMSProp~\cite{hinton2012neural} algorithm is used to update the network parameters with a learning rate of $2 \times 10^{-5}$. {`Ramp filter' is used for the filtering process in our method and both `Ramp' and `Hann' filters are adopted for other comparative FBP methods. In the following experiments, we will use FBP-R to denote FBP with the Ramp filter and FBP-H to denote FBP with the Hann filter.} The number of training epochs is set to 100. The batch size is set to 1, while the momentum and weight decay are set to 0.9 and 0.0, respectively. The number of basis functions is set to be 3 and 5 for the F-\ourmodel{} and L-\ourmodel{}, respectively. 
The mean square error (MSE) is adopted as the loss function for the proposed network, which minimizes the difference between the reconstructed image $I^*$ and the reference image $I$. The loss function is then given by
\begin{equation}\label{eq:loss_I}
    \mathcal{L}(I^*, I) = \Vert I^* - I \Vert^2_2.
\end{equation}
In this study, our network $\operatorname{Net}_w$ simply consists of a stack of two 1D convolutional layers and a ReLU function, in the form of `Conv. + ReLU + Conv.'. The following experiments reveal that even with such a simple network architecture, evident improvements in CT reconstruction can still be achieved.

The objective image analysis, \textit{i.e.}, peak signal-to-noise ratio (PSNR), feature similarity indexing method (FSIM), normalized mean square error (NMSE), modulation transfer function (MTF), and noise power spectrum (NPS), are used to evaluate the reconstruction quality. And subjective image analysis is also utilized. Subjective image noise, sharpness, contrast, diagnostic confidence and artifacts are evaluated independently by five radiologists with experience in CT examinations using a five-point scale (\textit{i.e.}, 1 = Poor, 2 = Limited, 3 = Average, 4 = Good, 5 = Excellent). Moreover, five different methods are adopted for comparison. {The FBP algorithm with nearest interpolation strategy (Ne FBP),  the FBP algorithm with cubic interpolation strategy (Cu FBP), and the FBP algorithm with linear interpolation strategy (Li FBP) are selected.} In addition, the deep learning based CT reconstruction network iRadonMAP~\cite{he2020radon}, and image-based DL network DICDNet~\cite{wang2021dicdnet} are also chosen. It should be noted that the DICDNet is designed for metal artifact reduction in the image domain and also proven to be able to finely remove noise-induced artifacts in the CT images. The related network settings are determined according to the suggestions from the original literatures.


\section{Experimental Results}\label{sec:results}

\renewcommand{\tabcolsep}{2 pt}{
\begin{figure}[t!]
	\begin{center}
		\begin{tabular}{ccc}
			
			\includegraphics[width=0.32\linewidth]{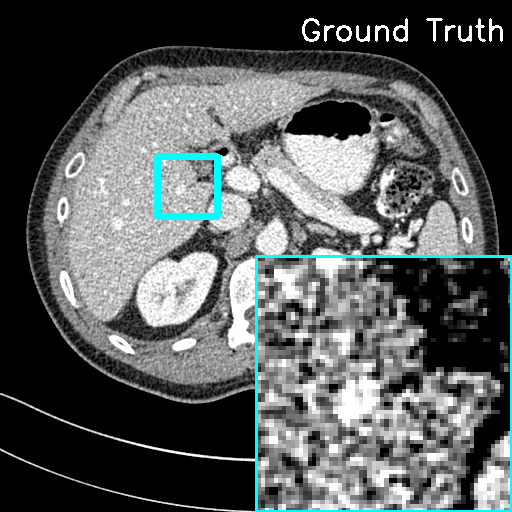}  & \includegraphics[width=0.32\linewidth]{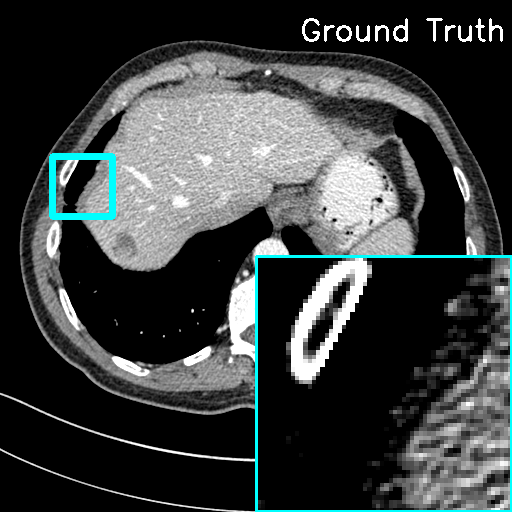}  & \includegraphics[width=0.32\linewidth]{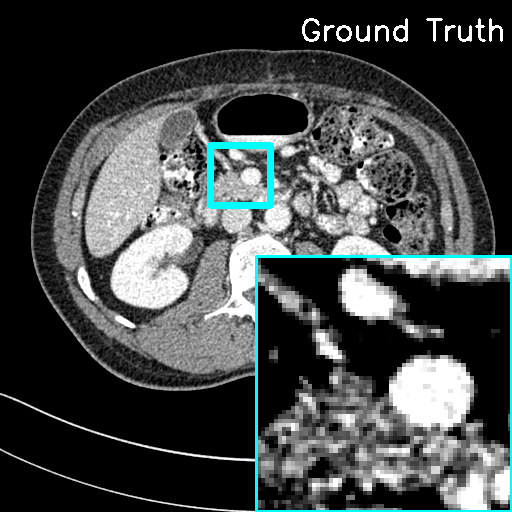} \\
   
			\includegraphics[width=0.32\linewidth]{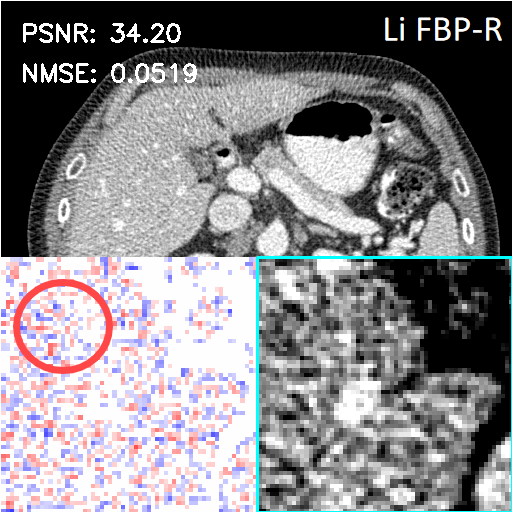}  & \includegraphics[width=0.32\linewidth]{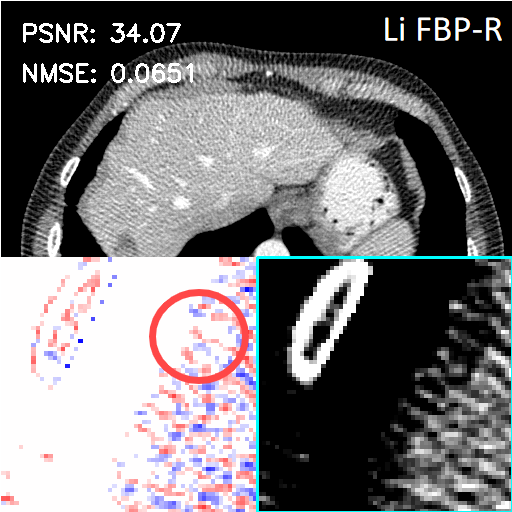}  & \includegraphics[width=0.32\linewidth]{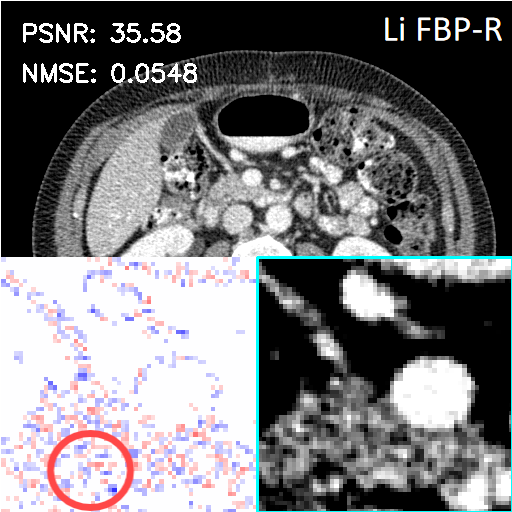} \\

   \includegraphics[width=0.32\linewidth]{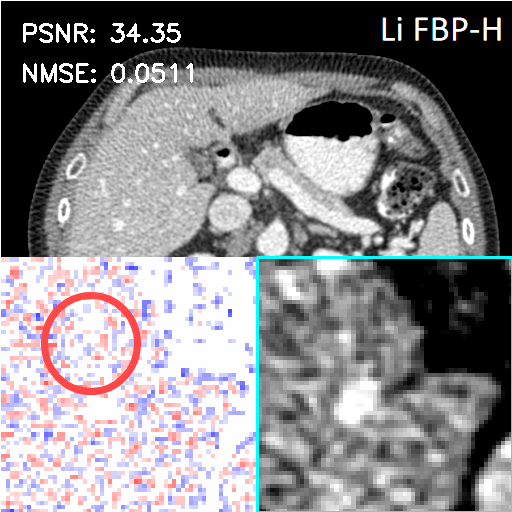}  & \includegraphics[width=0.32\linewidth]{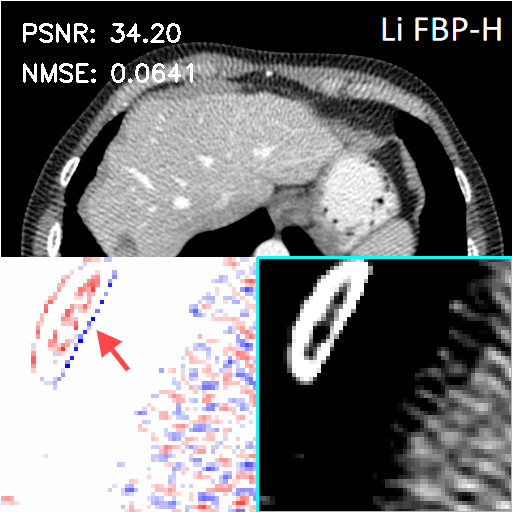}  & \includegraphics[width=0.32\linewidth]{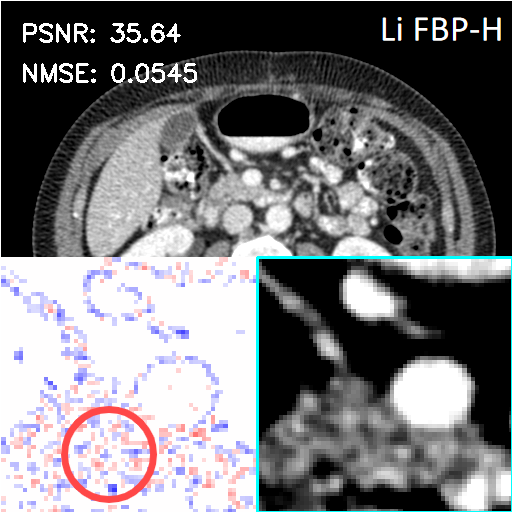} \\
   
			\includegraphics[width=0.32\linewidth]{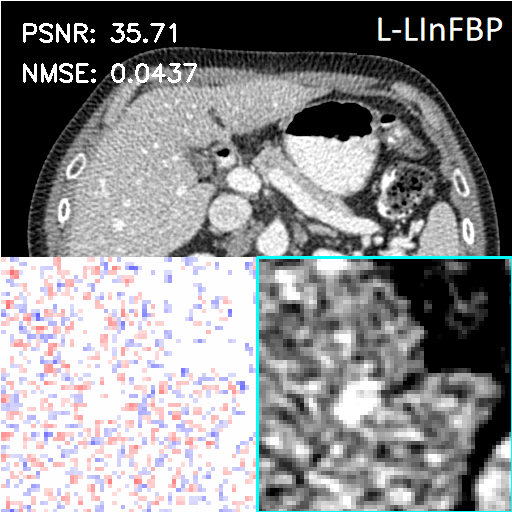}  & \includegraphics[width=0.32\linewidth]{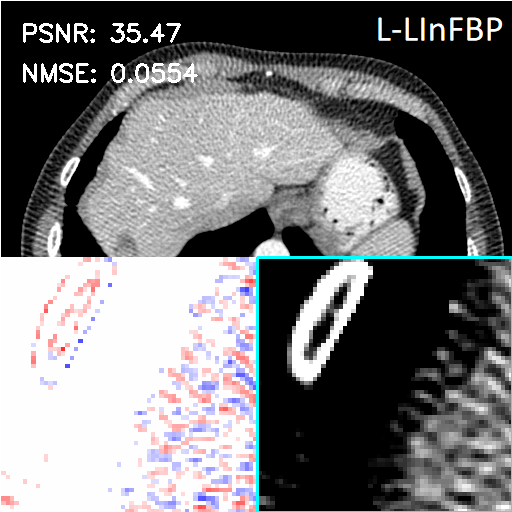}  & \includegraphics[width=0.32\linewidth]{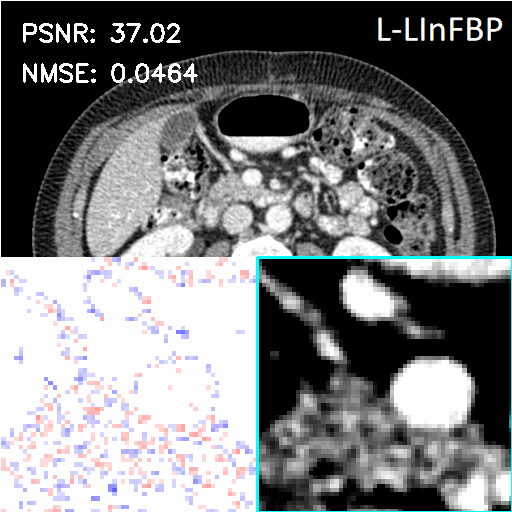} \\

             \includegraphics[width=0.32\linewidth]{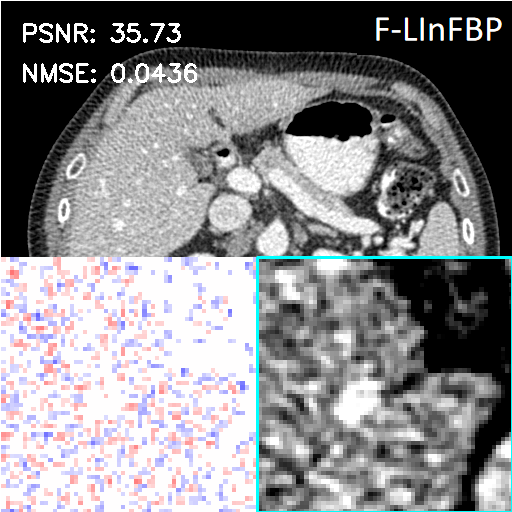}  & \includegraphics[width=0.32\linewidth]{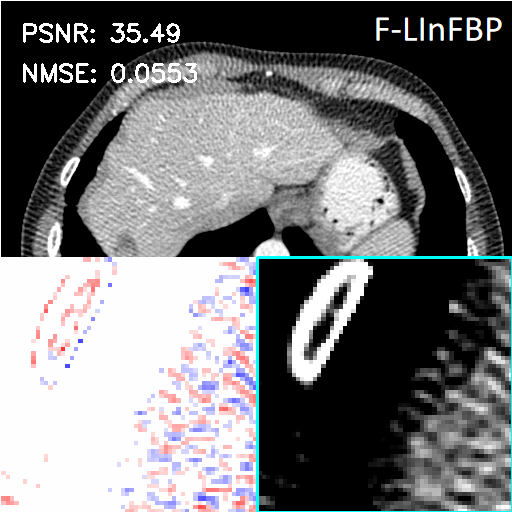}  & \includegraphics[width=0.32\linewidth]{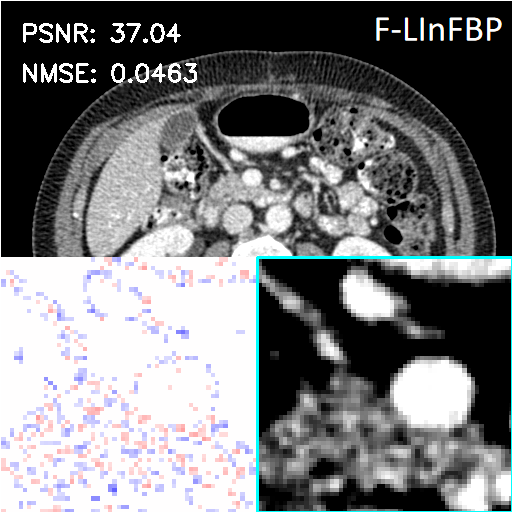} \\

		\end{tabular}	
		\caption{{Visual comparison of the representative images at only quarter-view case. Zoomed-in ROIs and error images indicated by the cyan boxes and and the quantitative assessments are illustrated on the corresponding images. The display window is [-160,240] HU and the zoom-in window is [60,200] HU.}}
		\label{fig:normaldose}
	\end{center}\vspace{-4mm}
\end{figure}}

\subsection{Results on full-view at normal- and quarter-dose cases}
To demonstrate the feasibility of the proposed \ourmodel{}, the 720 full view sinograms at normal- and quarter-dose levels are reconstructed {by the FBP with nearest and linear interpolation strategy (Ne FBP and Li FBP)}, FBP with Fourier \ourmodel{} (F-\ourmodel{}) and FBP with Linear \ourmodel{} (L-\ourmodel{}). And then the reconstructed results are evaluated quantitatively compared with the ground-truth. The comparison metrics are presented in Table~\ref{tab:720_views}. Wilcoxon signed-rank test results show that $p \ll 0.001$ for all evaluation metrics achieved by F-\ourmodel{}, L-\ourmodel{} and FBP, indicating a statistically great improvement of the proposed methods over both the Nearest FBP and Linear FBP. \HUI{Although the FSIM metric for Li FBP-H under low-dose conditions is slightly higher than that of our proposed methods, it is important to note that learning-based approaches inherently tend to generate smoother reconstructions, which may lead to a modest reduction in FSIM. Nevertheless, our methods achieve substantial improvements in PSNR and NMSE, indicating a significant enhancement in overall reconstruction quality.}

\renewcommand{\tabcolsep}{2 pt}{
\begin{figure}[t!]
	\begin{center}
		\begin{tabular}{ccc}
			
			\includegraphics[width=0.32\linewidth]{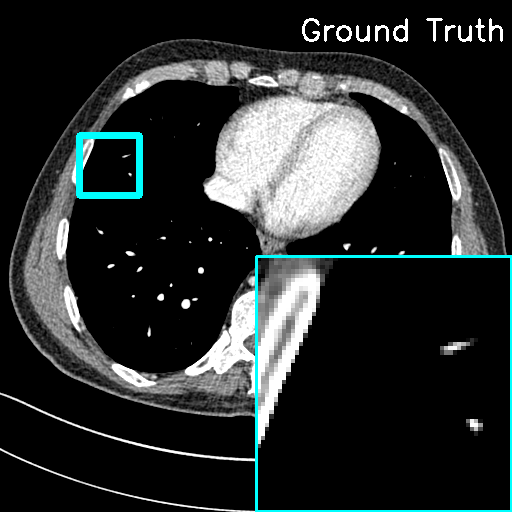}  & \includegraphics[width=0.32\linewidth]{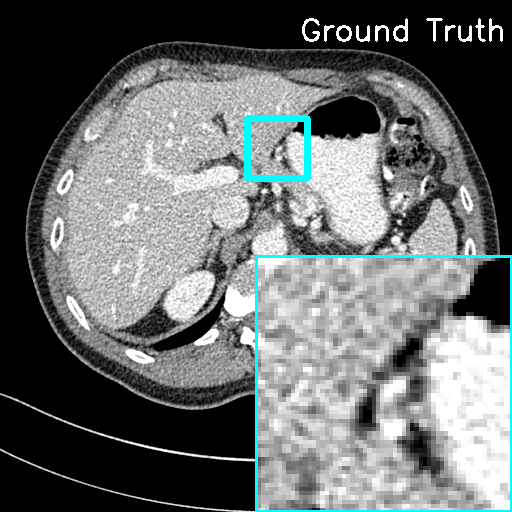}  & \includegraphics[width=0.32\linewidth]{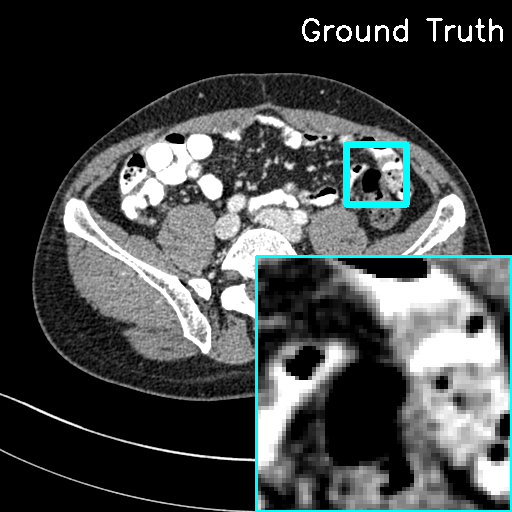} \\
   
			\includegraphics[width=0.32\linewidth]{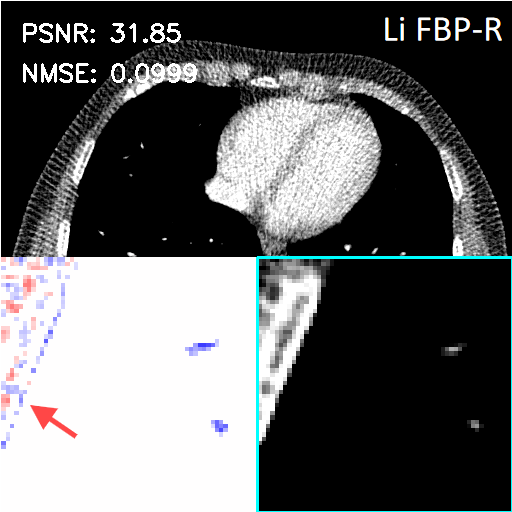}  & \includegraphics[width=0.32\linewidth]{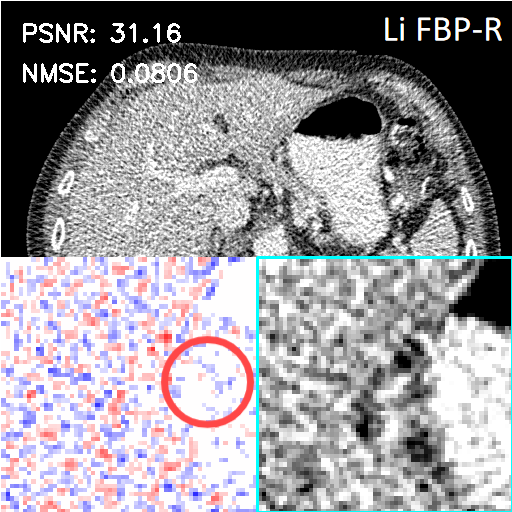}  & \includegraphics[width=0.32\linewidth]{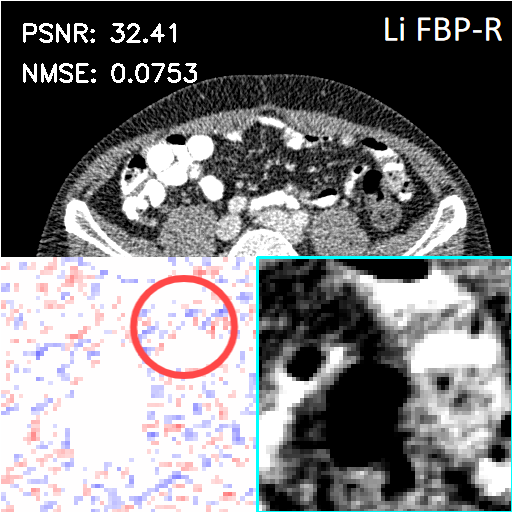} \\

   \includegraphics[width=0.32\linewidth]{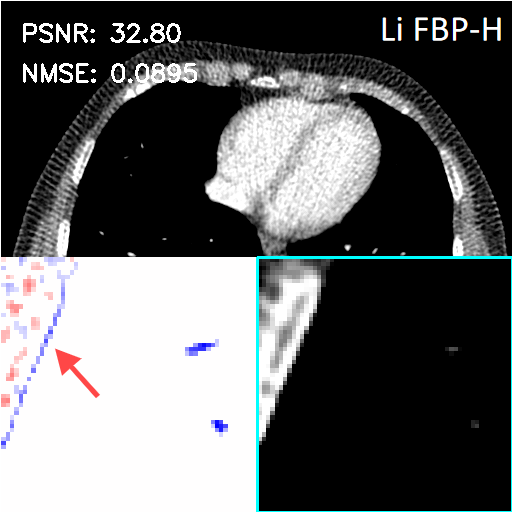}  & \includegraphics[width=0.32\linewidth]{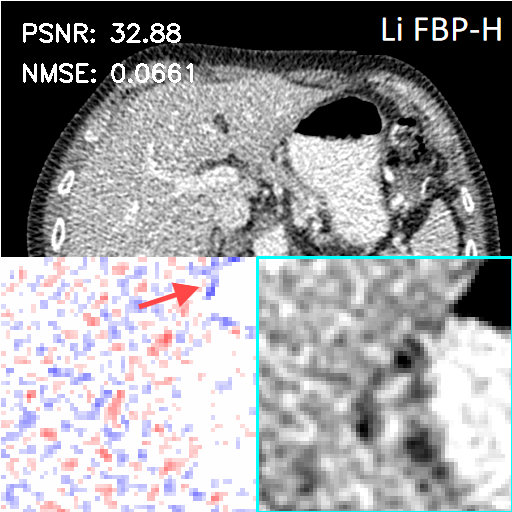}  & \includegraphics[width=0.32\linewidth]{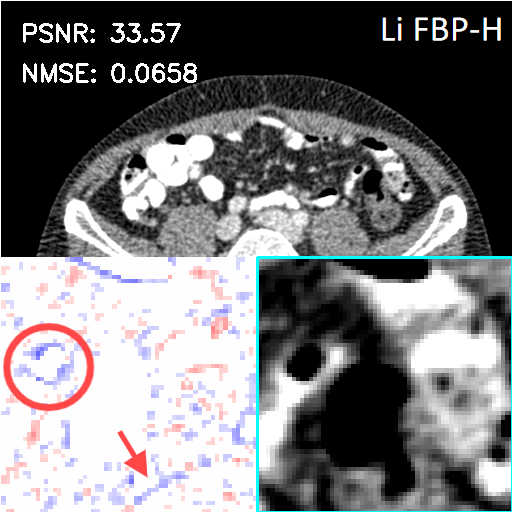} \\
   
			\includegraphics[width=0.32\linewidth]{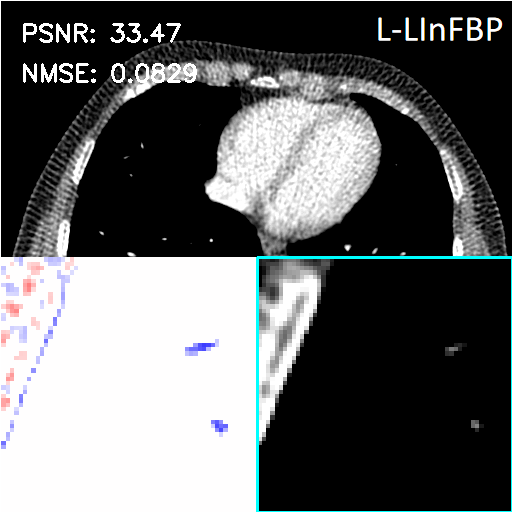}  & \includegraphics[width=0.32\linewidth]{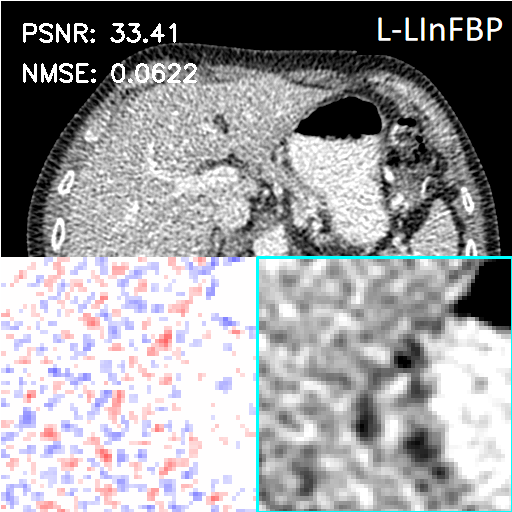}  & \includegraphics[width=0.32\linewidth]{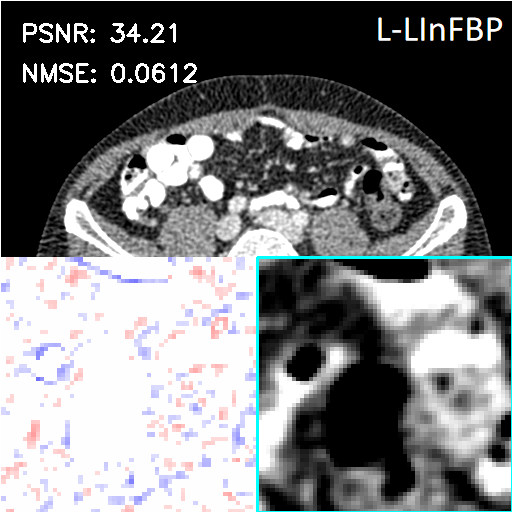} \\

             \includegraphics[width=0.32\linewidth]{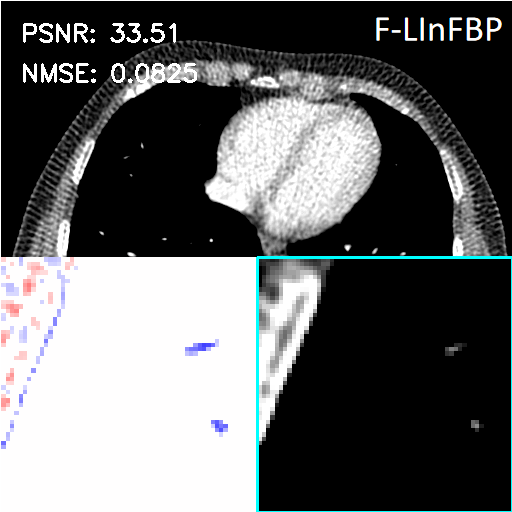}  & \includegraphics[width=0.32\linewidth]{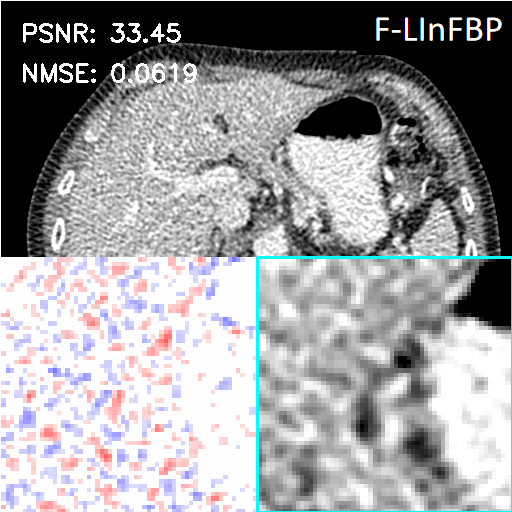}  & \includegraphics[width=0.32\linewidth]{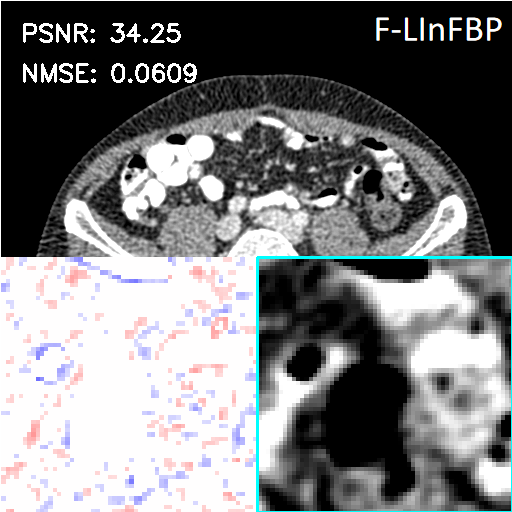} \\

		\end{tabular}
		\caption{{Visual comparison of the representative images at quarter-view with quarter-dose case. Zoomed-in ROIs and error images are indicated by the green boxes and and the quantitative assessments are illustrated on the corresponding images. The display window is [-160,240] HU and the zoom-in window is [-100,200] HU.}}
		\label{fig:lowdose}
	\end{center}
\vspace{-4mm}
\end{figure}}

\subsection{Results on quarter-view cases}

We assess the reconstruction performance of \ourmodel{} under quarter-view cases, i.e., only quarter-view and quarter-view with quarter-dose conditions. Fig.~\ref{fig:normaldose} and Fig.~\ref{fig:lowdose} present the reconstructed CT images using Li FBP-R, Li FBP-H and our methods at quarter-view cases and quarter-view with quarter-dose cases, respectively. It can be observed from the visualizations that the proposed methods yield better results closer to the ground-truth with higher PSNR scores at all the cases, as indicated by the red circles and arrows in the error maps, compared to Linear FBP methods. Specifically, in Fig.~\ref{fig:lowdose}, both F-\ourmodel{} and L-\ourmodel{} can reconstruct highly accurate structure textures with negligible artifacts compared to the ground-truth. The improvements in image quality can be ascribed to the accurate radon transform interpolation obtained by the \ourmodel{}, which implies its effectiveness in low-dose CT imaging applications while maintaining image quality. It should also be noted that the L-\ourmodel{} produces almost similar reconstruction performance with F-\ourmodel{}, as seen in the least absolute errors compared to the ground-truth. This indicates that the linear basis functions in L-\ourmodel{} can already represent the continuous backprojection in satisfactory accuracy while with evidently less complexity as compared to F-\ourmodel{}. 

For quantitative evaluation,  Table~\ref{tab:290_views} presents the quantitative measurements of the CT images reconstructed by the FBP algorithms with different interpolation strategies and the F-\ourmodel{} and L-\ourmodel{} methods at only quarter-view and quarter-view with quarter-dose conditions, respectively. It is seen that both the proposed F-\ourmodel{} and L-\ourmodel{} consistently achieve higher PSNR scores, lower NMSE scores, and higher FSIM scores for all the quarter-view cases and quarter-view with quarter-dose cases compared to other competing methods. And the L-\ourmodel{} can obtain similar quantitative measurements with the F-\ourmodel{}, which is consistent with the visual inspection.

\def\arraystretch{1.1}
\renewcommand{\tabcolsep}{6 pt}{
\begin{table}
\caption{Quantitative measures of the CT reconstructions for different methods under normal- and quarter-dose conditions. }
\label{tab:720_views}
\small
\begin{center}
\begin{tabular}{lccc}
\toprule[1pt]
Method & PSNR & NMSE & FSIM \\
\midrule
\multicolumn{4}{c}{Normal dose}\\
\midrule
 {Ne FBP-R} & 39.296$\pm$0.89 & 0.0355$\pm$0.004 & 0.9934$\pm$0.0010 \\
 {Li FBP-R} & 40.593$\pm$0.90 & 0.0306$\pm$0.003 & 0.9944$\pm$0.0008 \\
 {Li FBP-H} & 37.479$\pm$0.93 & 0.0439$\pm$0.005 & 0.9865$\pm$0.0020 \\
 F-\ourmodel{} & 45.311$\pm$0.91 & 0.0178$\pm$0.002 & 0.9986$\pm$0.0002 \\
 L-\ourmodel{} & 44.273$\pm$0.90 & 0.0200$\pm$0.002 & 0.9982$\pm$0.0003 \\
 \midrule
\multicolumn{4}{c}{Low dose}\\
\midrule
 {Ne FBP-R}  & 35.246$\pm$1.06 & 0.0565$\pm$0.003 & 0.9693$\pm$0.006 \\
 {Li FBP-R} & 36.483$\pm$1.06 & 0.0490$\pm$0.003 & 0.9730$\pm$0.006 \\
 {Li FBP-H} & 35.945$\pm$0.91 & 0.0522$\pm$0.004 & 0.9751$\pm$0.004 \\
 F-\ourmodel{} & 37.514$\pm$1.13 & 0.0435$\pm$0.003 & 0.9712$\pm$0.006 \\
 L-\ourmodel{} & 37.294$\pm$1.02 & 0.0446$\pm$0.003 & 0.9720$\pm$0.006 \\
\bottomrule[1pt]
\end{tabular}
\end{center}
\end{table}}

\def\arraystretch{1.1}
\renewcommand{\tabcolsep}{6 pt}{
\begin{table}
\caption{Quantitative measures of the CT reconstructions for different methods under only quarter-view and quarter-view
with quarter-dose conditions.}
\label{tab:290_views}
\small
\begin{center}
\begin{tabular}{lccc}
\toprule[1pt]
Method & PSNR & NMSE & FSIM \\
\midrule
\multicolumn{4}{c}{Normal dose}\\
\midrule
 {Ne FBP-R} & 33.624$\pm$0.92 & 0.0684$\pm$0.008 & 0.9417$\pm$0.007 \\
 {Cu FBP-R} & 34.535$\pm$0.94 & 0.0616$\pm$0.008 & 0.9440$\pm$0.007 \\
 {Li FBP-R} & 35.004$\pm$0.93 & 0.0584$\pm$0.007 & 0.9463$\pm$0.006 \\
 {Li FBP-H} & 35.205$\pm$0.92 & 0.0570$\pm$0.007 & 0.9670$\pm$0.005 \\
 F-\ourmodel{} & 36.397$\pm$0.94 & 0.0498$\pm$0.006 & 0.9702$\pm$0.005 \\
 L-\ourmodel{} & 36.337$\pm$0.94 & 0.0501$\pm$0.006 & 0.9688$\pm$0.005 \\
\midrule
\multicolumn{4}{c}{Low dose}\\
\midrule
 {Ne FBP-R} & 31.036$\pm$0.94 & 0.0918$\pm$0.007 & 0.9293$\pm$0.009 \\
 {Cu FBP-R} & 31.678$\pm$0.96 & 0.0853$\pm$0.006 & 0.9295$\pm$0.009 \\
 {Li FBP-R} & 32.529$\pm$0.93 & 0.0773$\pm$0.006 & 0.9368$\pm$0.008 \\
 {Li FBP-H} & 33.904$\pm$0.88 & 0.0661$\pm$0.006 & 0.9502$\pm$0.006 \\
 F-\ourmodel{} & 34.555$\pm$0.89 & 0.0613$\pm$0.005 & 0.9510$\pm$0.006 \\
 L-\ourmodel{} & 34.549$\pm$0.89 & 0.0613$\pm$0.005 & 0.9520$\pm$0.006 \\
\bottomrule[1pt]
\end{tabular}
\end{center}
\end{table}}

\subsection{Results on more sparse-view cases}

{To assess the robustness of the proposed \ourmodel{}, the data at 360 views, 145 views, 100 views, 72 views and 36 views are utilized.} Fig.~\ref{fig:100views} shows the CT images reconstructed by Li FBP-R, Li FBP-H, F-\ourmodel{} and L-\ourmodel{} methods at 100 views. It can be seen that the proposed F-\ourmodel{} and L-\ourmodel{} can properly learn the continuous backprojection representation, leading to the effective recovery of some fine structures compared to the direct FBP-reconstructed CT images. 
Furthermore, Table~\ref{tab:100_sparse_view} presents a comparison of evaluation metrics among the FBP, F-\ourmodel{}, and L-\ourmodel{} methods based on tests with three different sampling views. As the number of sampling views increases, the quantitative assessments of all competing methods achieve gradual improvements in reconstruction performance. Notably, both the proposed F-\ourmodel{} and L-\ourmodel{} methods outperform the traditional FBP algorithm. And interestingly, despite its simpler basis function constructions, the proposed L-\ourmodel{} demonstrates the potential to achieve comparable or even better reconstruction performances compared to the proposed F-\ourmodel{} with more complex Fourier basis functions, as also illustrated in Fig.~\ref{fig:100views}.


\def\arraystretch{1.1}
\renewcommand{\tabcolsep}{7 pt}{
\begin{table}
\caption{{Quantitative measures of the CT reconstructions for different methods under 36, 72, 100, 145 and 360 views conditions.}}
\label{tab:100_sparse_view}
\small
\begin{center}
\begin{tabular}{lccc}
\toprule[1pt]
Method & PSNR & NMSE & FSIM \\
\midrule
\multicolumn{4}{c}{{36 views}}\\
\midrule
{Ne FBP-R} & 19.639$\pm$0.94 & 0.3426$\pm$0.044 & 0.3426$\pm$0.044 \\
{Li FBP-R} & 20.106$\pm$0.94 & 0.3247$\pm$0.042 & 0.6365$\pm$0.021 \\
{Li FBP-H} & 21.043$\pm$0.94 & 0.2916$\pm$0.038 & 0.6673$\pm$0.021 \\
{F-\ourmodel{}} & 23.281$\pm$0.94 & 0.2252$\pm$0.028 & 0.7613$\pm$0.020 \\
{L-\ourmodel{}} & 23.113$\pm$0.93 & 0.2296$\pm$0.029 & 0.7504$\pm$0.020 \\
\midrule
\multicolumn{4}{c}{72 views}\\
\midrule
 {Ne FBP-R} & 23.656$\pm$0.85 & 0.2157$\pm$0.024 & 0.7435$\pm$0.017 \\
 {Li FBP-R} & 24.247$\pm$0.87 & 0.2015$\pm$0.026 & 0.7363$\pm$0.021 \\
 {Li FBP-H} & 25.325$\pm$0.85 & 0.1780$\pm$0.022 & 0.7638$\pm$0.020 \\
 F-\ourmodel{} & 26.962$\pm$0.82 & 0.1474$\pm$0.017 & 0.8234$\pm$0.016 \\
 L-\ourmodel{} & 26.916$\pm$0.82 & 0.1481$\pm$0.017 & 0.8180$\pm$0.016 \\
\midrule
\multicolumn{4}{c}{100 views}\\
\midrule
{Ne FBP-R} & 25.789$\pm$0.86 & 0.1687$\pm$0.018 & 0.7835$\pm$0.018 \\
 {Li FBP-R} & 26.462$\pm$0.89 & 0.1561$\pm$0.020 & 0.7903$\pm$0.020 \\
  {Li FBP-H} & 27.538$\pm$0.87 & 0.1379$\pm$0.017 & 0.8144$\pm$0.019 \\
 F-\ourmodel{} & 28.707$\pm$0.82 & 0.1205$\pm$0.013 & 0.8551$\pm$0.014 \\
 L-\ourmodel{} & 28.689$\pm$0.83 & 0.1208$\pm$0.013 & 0.8556$\pm$0.014 \\
\midrule
\multicolumn{4}{c}{145 views}\\
\midrule
 {Ne FBP-R} & 28.099$\pm$0.88 & 0.1292$\pm$0.013 & 0.8447$\pm$0.015 \\
 {Li FBP-R} & 28.911$\pm$0.87 & 0.1178$\pm$0.014 & 0.8509$\pm$0.016 \\
 {Li FBP-H} & 30.011$\pm$0.85 & 0.1037$\pm$0.012 & 0.8701$\pm$0.014 \\
 F-\ourmodel{} & 30.804$\pm$0.82 & 0.0946$\pm$0.010 & 0.8915$\pm$0.012 \\
 L-\ourmodel{} & 30.802$\pm$0.82 & 0.0946$\pm$0.010 & 0.8903$\pm$0.012 \\
\midrule
\multicolumn{4}{c}{{360 views}}\\
\midrule
{Ne FBP-R} & 35.496$\pm$0.93 & 0.0551$\pm$0.007 & 0.9798$\pm$0.004 \\	
{Li FBP-R} & 36.981$\pm$0.94 & 0.0465$\pm$0.006 & 0.9831$\pm$0.003 \\
{Li FBP-H} & 36.339$\pm$0.92 & 0.0501$\pm$0.006 & 0.9802$\pm$0.030 \\
{F-\ourmodel{}} & 38.526$\pm$0.95 & 0.0389$\pm$0.005 & 0.9864$\pm$0.002 \\
{L-\ourmodel{}} & 38.314$\pm$0.95 & 0.0399$\pm$0.005 & 0.9864$\pm$0.003 \\
\bottomrule[1pt]
\end{tabular}
\end{center}
\end{table}}

\renewcommand{\tabcolsep}{2 pt}{
\begin{figure}[t!]
	\begin{center}
		\begin{tabular}{ccc}
			
			\includegraphics[width=0.32\linewidth]{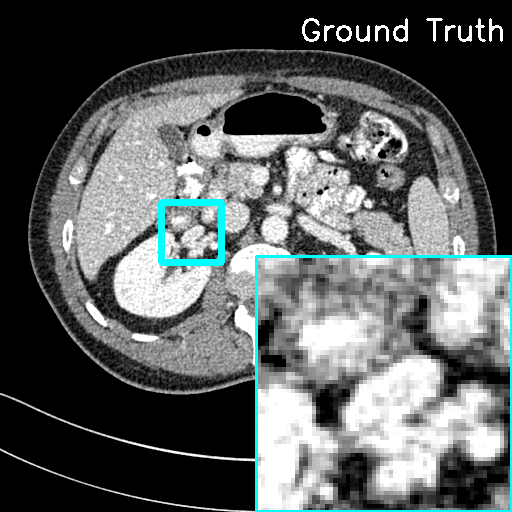}  & \includegraphics[width=0.32\linewidth]{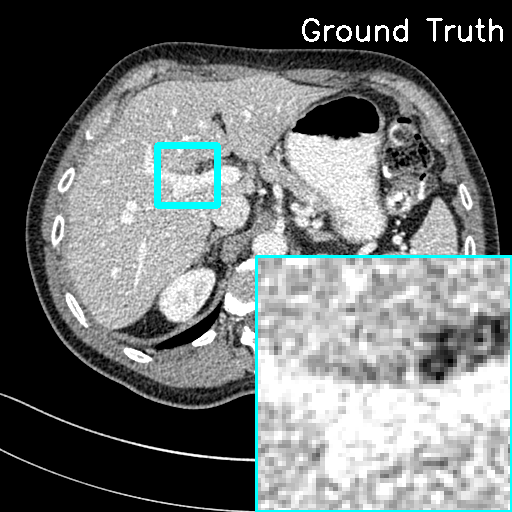}  & \includegraphics[width=0.32\linewidth]{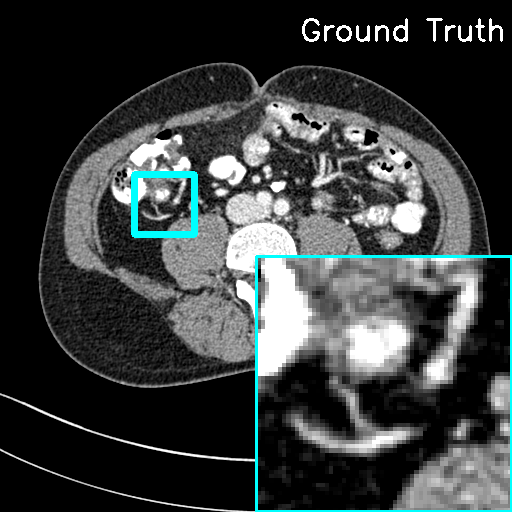} \\
   
			\includegraphics[width=0.32\linewidth]{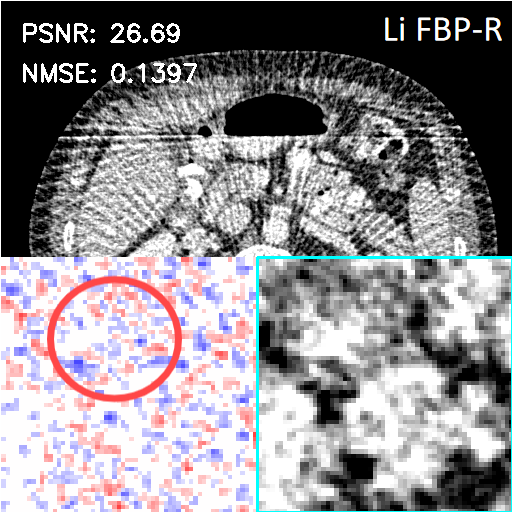}  & \includegraphics[width=0.32\linewidth]{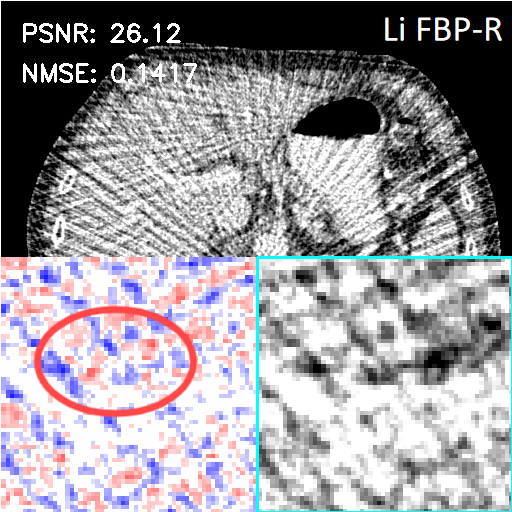}  & \includegraphics[width=0.32\linewidth]{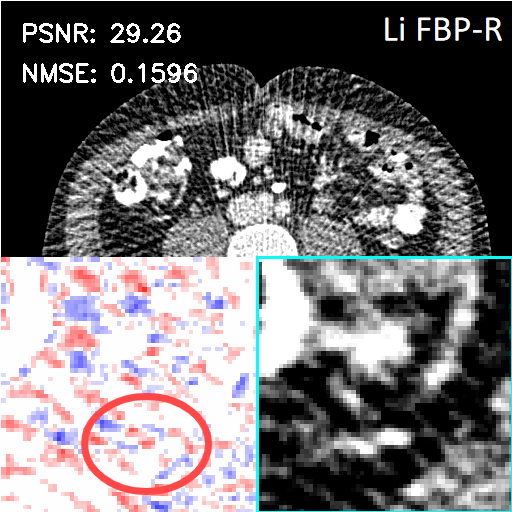} \\

            \includegraphics[width=0.32\linewidth]{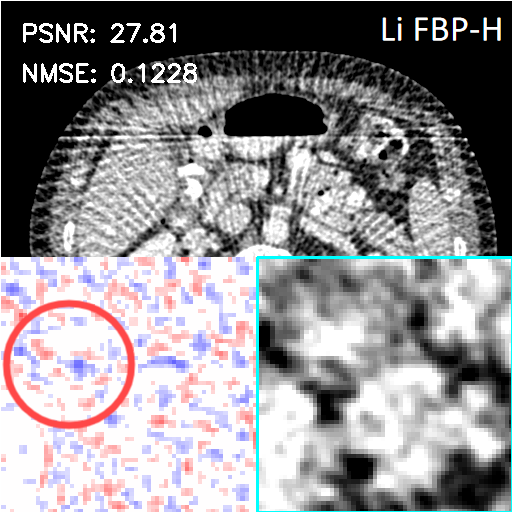}  & \includegraphics[width=0.32\linewidth]{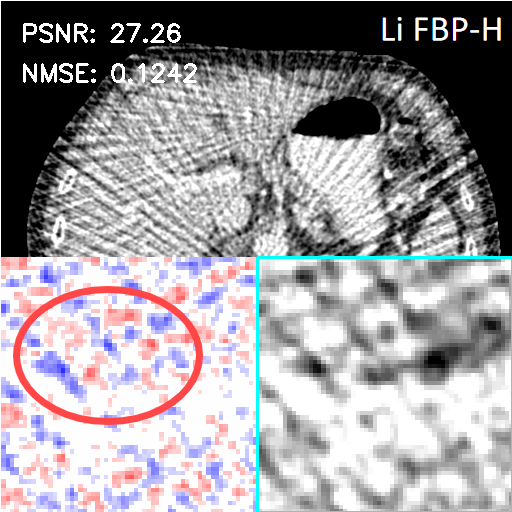}  & \includegraphics[width=0.32\linewidth]{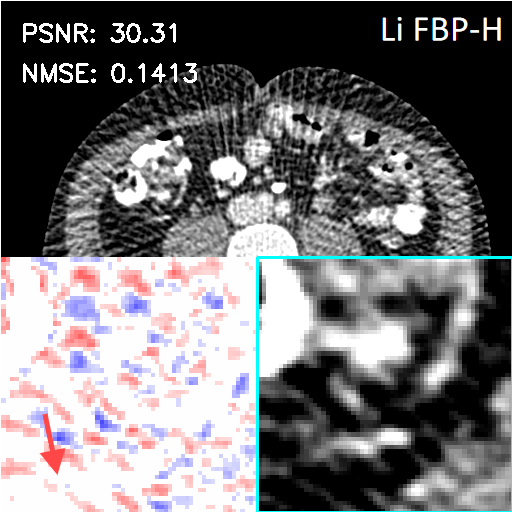} \\
   
			\includegraphics[width=0.32\linewidth]{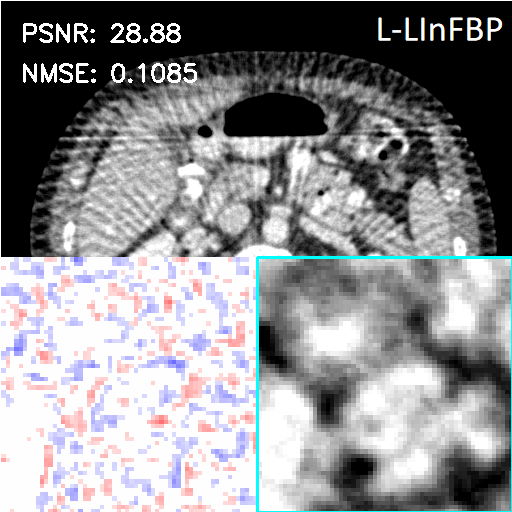}  & \includegraphics[width=0.32\linewidth]{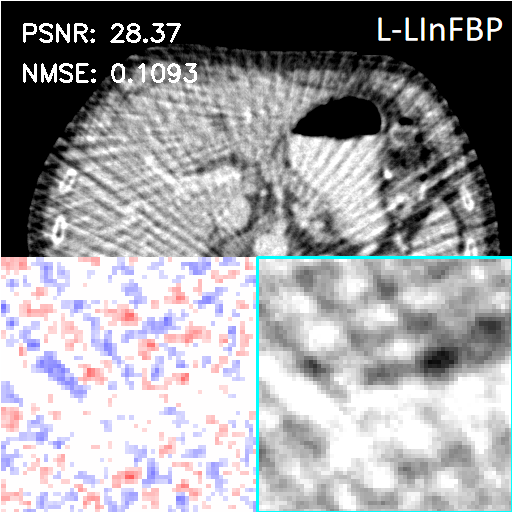}  & \includegraphics[width=0.32\linewidth]{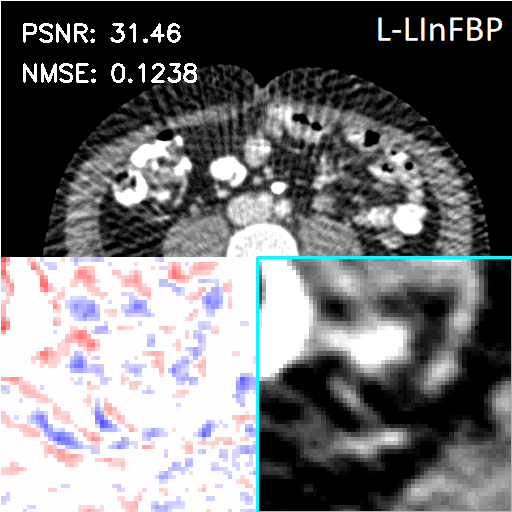} \\

             \includegraphics[width=0.32\linewidth]{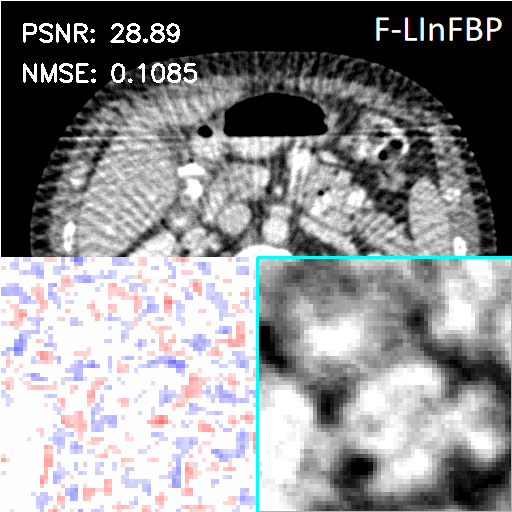}  & \includegraphics[width=0.32\linewidth]{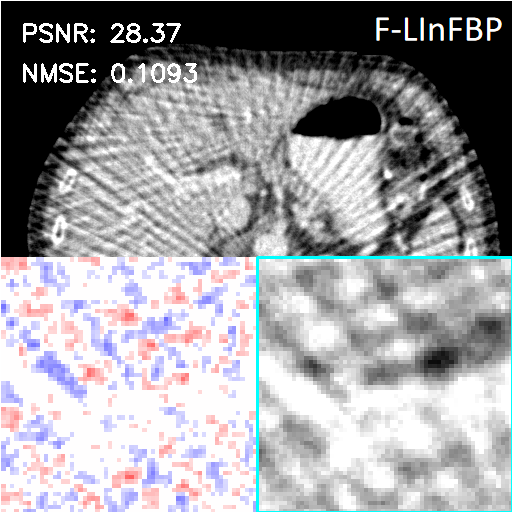}  & \includegraphics[width=0.32\linewidth]{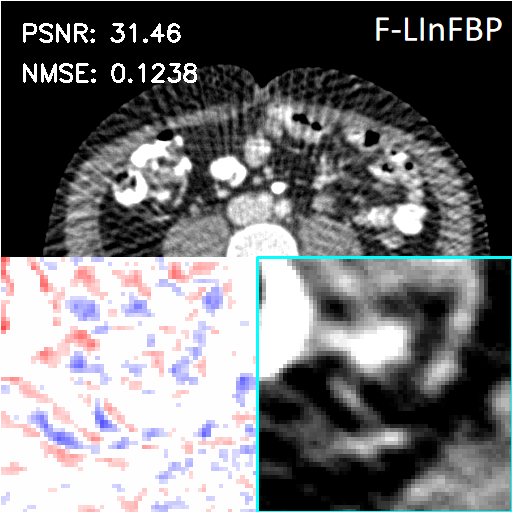} \\

		\end{tabular}
		\caption{{Visual comparison of the representative images at 100 views. Zoomed-in ROIs and error images  indicated by the green boxes and and the quantitative assessments are illustrated on the corresponding images. The display window is [-160,240] HU and the zoom-in window is [-100,200] HU.}}
		\label{fig:100views}
	\end{center}\vspace{-3mm}
\end{figure}}

\renewcommand{\tabcolsep}{3 pt}{
\begin{figure}[t!]
	\begin{center}
		\begin{tabular}{cc}
			
			\includegraphics[width=0.45\linewidth]{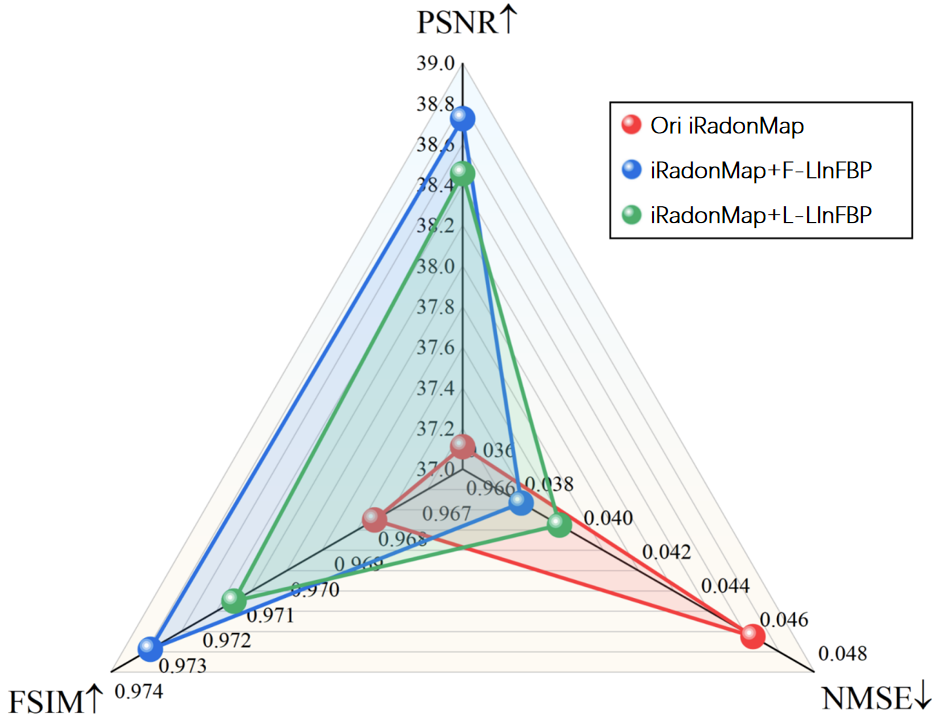}  &
			\includegraphics[width=0.45\linewidth]{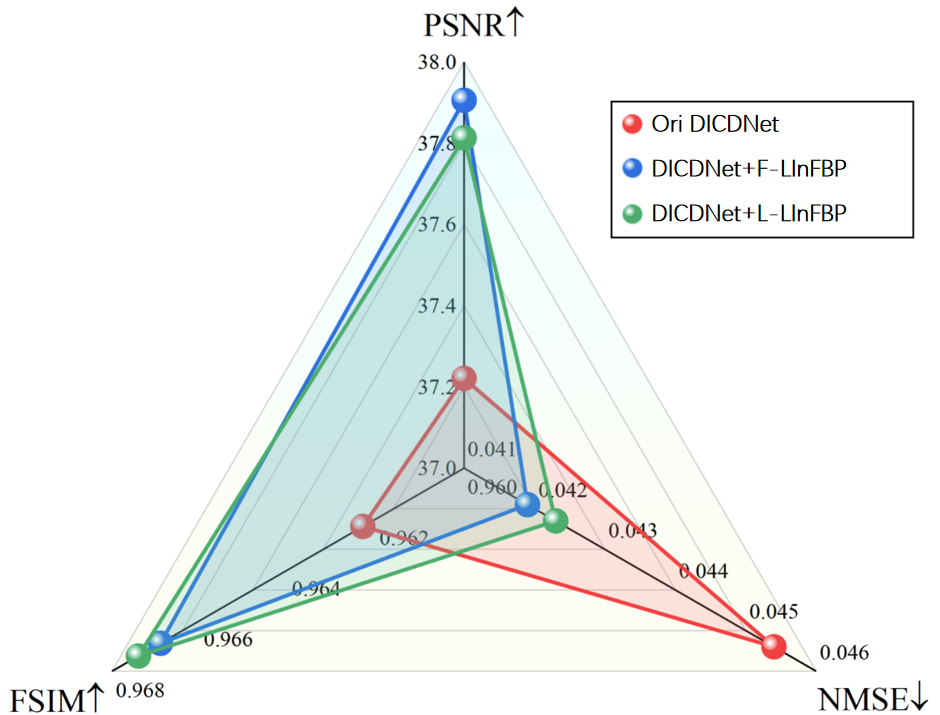}\\
   \includegraphics[width=0.45\linewidth]{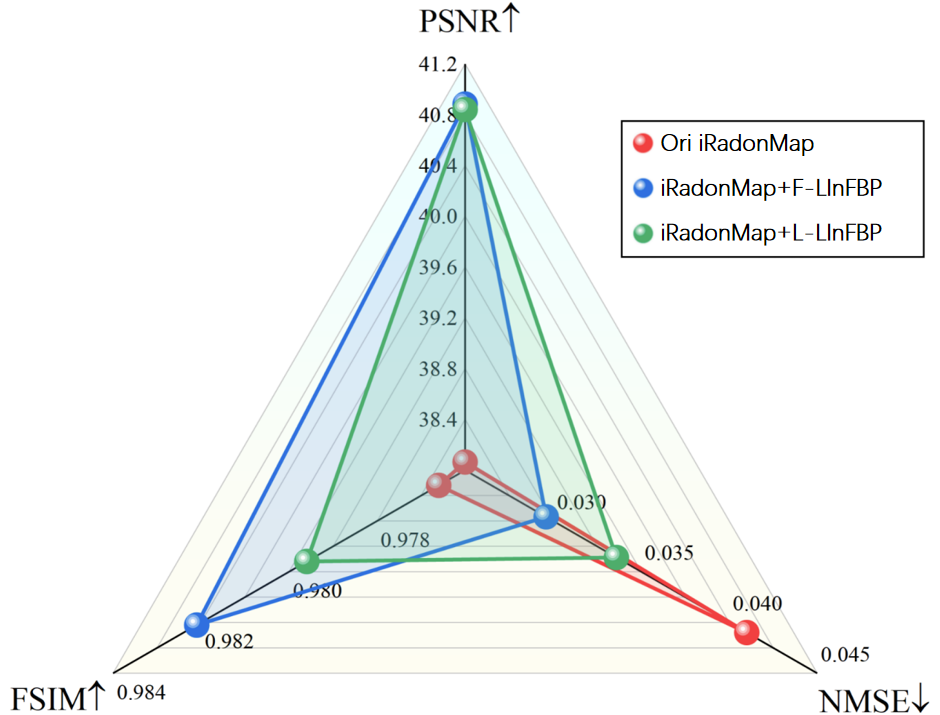}  &
			\includegraphics[width=0.45\linewidth]{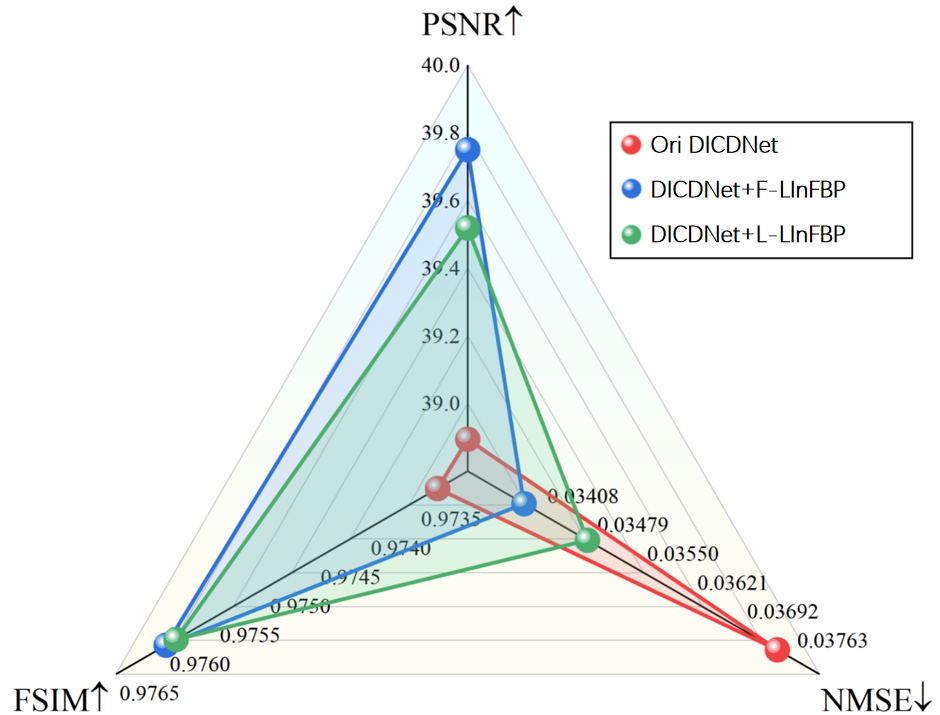}\\
   
		\end{tabular}
		\caption{The radar charts which display the quantitative measurements on selective reconstructed images from different methods at 100 views (first row) and 146 views (second row).}
		\label{fig:psnr}
	\end{center}
\end{figure}}

\subsection{Combination with State-of-the-art Methods}\label{sec:sota}

We then evaluate the `plug-and-play' capability of \ourmodel{} by integrating it into recent state-of-the-art CT reconstruction methods, including iRadonMap~\cite{he2020radon} and DICDNet~\cite{wang2021dicdnet}. Specifically, we replace the sinusoidal backprojection layer in iRadonMap with our continuous backprojection (\ourmodel{}) to form iRadonMap+F-\ourmodel{} and iRadonMap+L-\ourmodel{}. Similarly, we replace the traditional FBP module of DICDNet with \ourmodel{}. Fig.~\ref{fig:psnr} shows the radar charts which display the quantitative measurements on selective reconstructed images at 100 views (first row) and 146 views (second row) from iRadonMap, iRadonMap+F-\ourmodel{}, iRadonMap+L-\ourmodel{}, DICDNet, DICDNet+F-FBP and DICDNet+L-FBP, respectively. It is evident that the proposed \ourmodel{} consistently enhances the reconstruction performance of each model it is combined with. This improvement can be attributed to the effective enhancement of the accuracy of the radon transform by the proposed \ourmodel{}, enabling it to be seamlessly integrated into existing reconstruction methods to readily enhance their reconstruction performance. Fig.~\ref{fig:iRadon146} shows the reconstructed images and error maps obtained using different methods. When combined with the proposed \ourmodel{}, iRadonMap and DICDNet demonstrate superior performance compared to their counterparts without being equipped with \ourmodel{}, exhibiting fewer noise-induced artifacts and reduced residual errors in the error maps, as observed in the zoomed-in ROI images indicated by the cyan box.

{Moreover, Fig.~\ref{fig:nps} depicts the MTF curves of different methods, which refers to an image's ability to preserve contrast as resolution increases, in the ROI indicated by the red circle. \HUI{The calculation procedure of MTF is followed the works~\cite{zeng2015spectral, wu2021theory}.} It can be seen that the state-of-the-art approaches achieve an improvement when being integrated with our method. This improvement can be attributed to the design of the original iRadonMAP and DICDNet, which leverage multi-layer neural networks with robust denoising capabilities to ensure high-quality output with perhaps some degree of contrast smoothing. On the contrary, our method mitigates the interpolation errors in the backprojection process of these models, allowing for better preservation of image details. Consequently, our approach retains contrast more effectively, leading to higher MTF curves and overall improved performance. 
Specifically, the MTF curves of DICDNet+F-\ourmodel{} and DICDNet+L-\ourmodel{} overlap, indicating that F-\ourmodel{} and L-\ourmodel{} achieve similar contrast preservation in the images. These results demonstrate the effective `plug-and-play' capability of the proposed \ourmodel{}.}

\renewcommand{\tabcolsep}{2 pt}{
\begin{figure}[t!]
	\begin{center}
		\begin{tabular}{c}
		\includegraphics[width=0.98\linewidth]{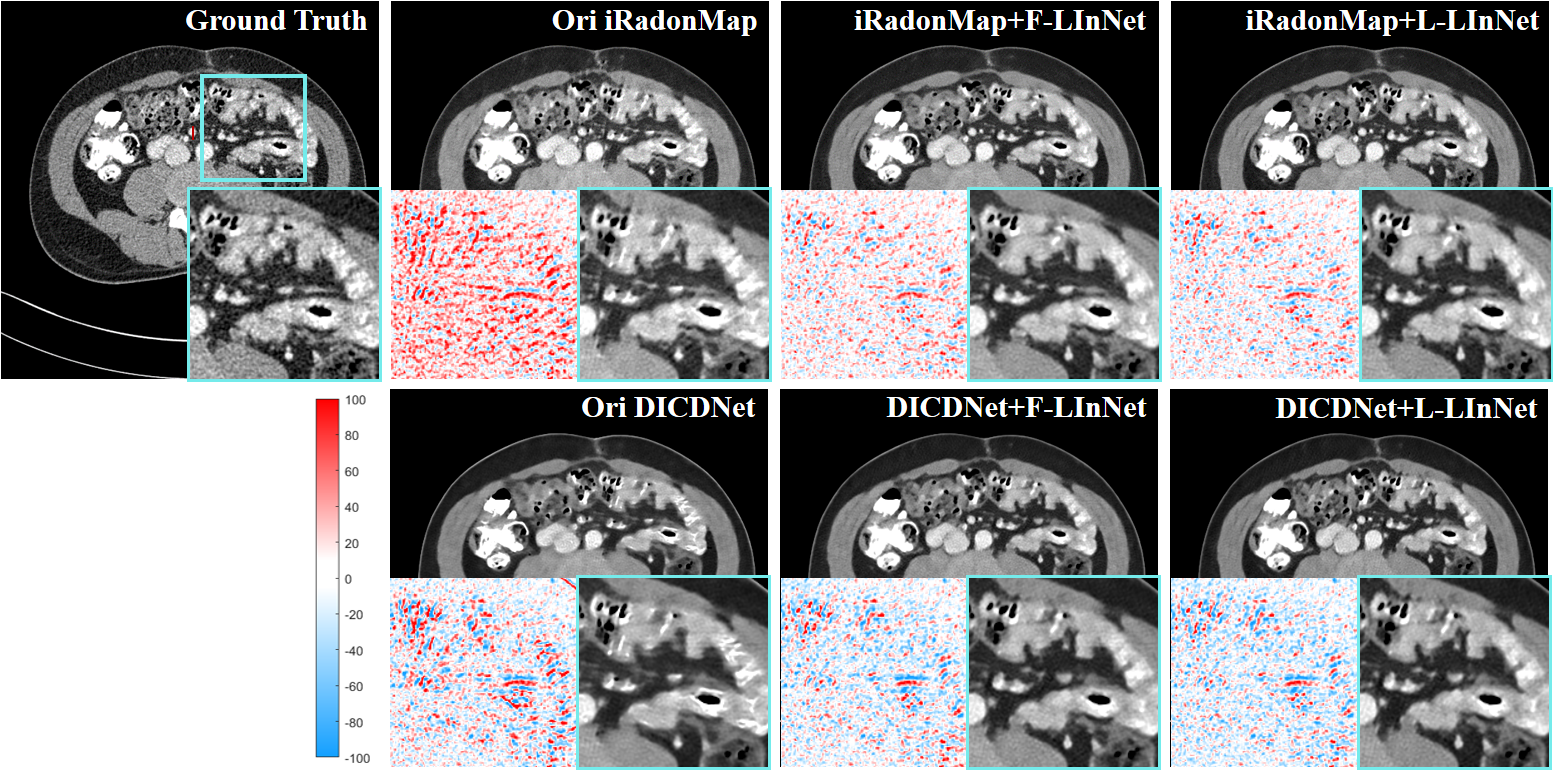} \\
        
        \end{tabular}
		\caption{Visual comparison of the representative images of the different methods at 146 views. Zoomed-in ROIs and error images indicated by the green boxes are illustrated on the corresponding images. The display window and the zoom-in window are [-160,240] HU.}
		\label{fig:iRadon146}
	\end{center}
 \vspace{-4mm}
\end{figure}}

\def\arraystretch{1.1}
\renewcommand{\tabcolsep}{2 pt}{
\begin{table}[t!]
\caption{Comparisons of training and testing geometries.}
\label{tab:geometries}
\small
\begin{center}
\begin{tabular}{cccccc}
\toprule[1pt]
Geometry & view & image pixel& pixel space (mm) & detector & bin (mm) \\
\midrule
Training & 1152 & 512$\times$512 & 0.6641$\times$0.6641 & 736 & 1.3696\\
Testing & 1152 & 512$\times$512 & 0.7422$\times$0.7422 & 736 & 1.3740\\
\bottomrule[1pt]
\end{tabular}\vspace{-2mm}
\end{center}
\end{table}}

\renewcommand{\tabcolsep}{3 pt}{
\begin{figure}[t!]
	\begin{center}
 \includegraphics[width=1\linewidth]{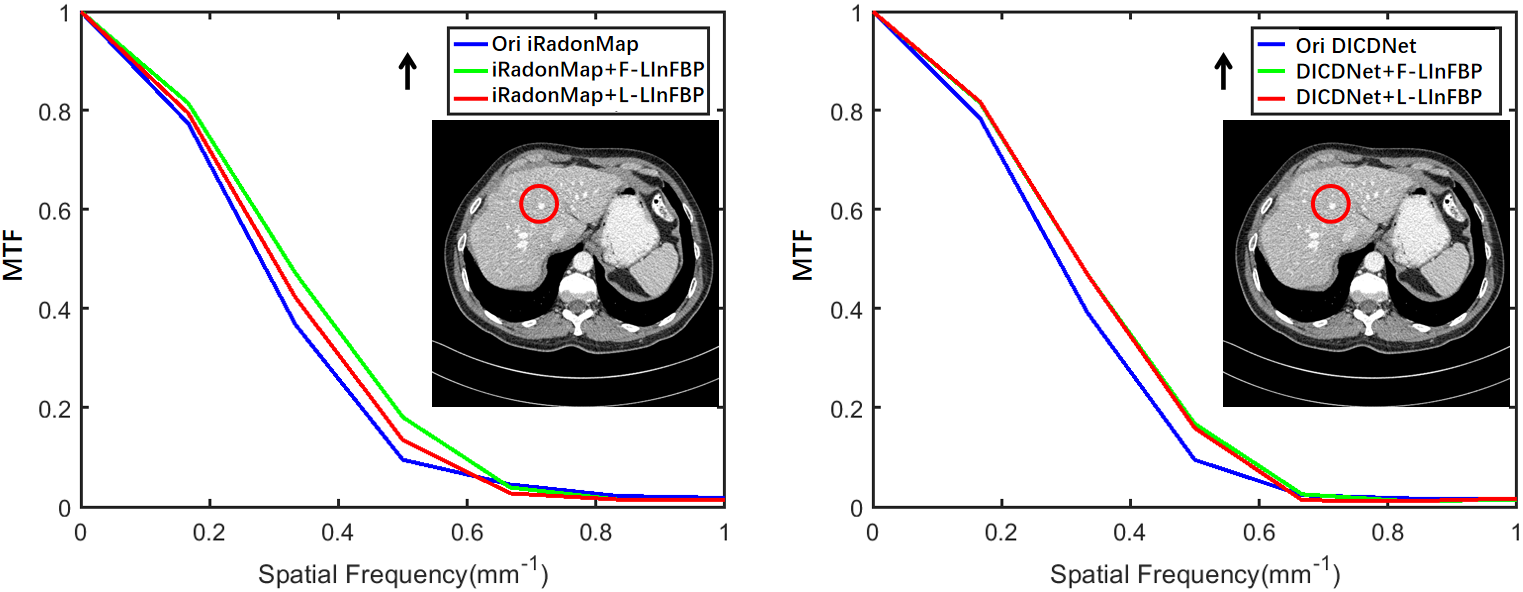}
		\caption{MTF curves from the region of interest (ROI) indicated by the red circle, in the reconstructed images obtained from various methods under 146 views. The curves demonstrate that our approach can help retain contrast and its effective `plug-and-play' capability.}
		\label{fig:nps}
	\end{center}
 \vspace{-2mm}
\end{figure}}

\def\arraystretch{1.1}
\renewcommand{\tabcolsep}{12 pt}{
\begin{table}
\caption{Quantitative measures of CT reconstructions for different methods under different geometries.}
\label{tab:different}
\small
\begin{center}
\begin{tabular}{lccc}
\toprule[1pt]
Method & PSNR & NMSE & FSIM \\
\midrule
\multicolumn{4}{c}{Original Geometry}\\
\midrule
 Ne FBP-R & 40.617 & 0.0309 & 0.9946 \\
 Li FBP-R & 41.227 & 0.0285 & 0.9948 \\
 Li FBP-H & 37.532 & 0.0436 & 0.9866 \\
 DSigNet~\cite{he2021downsampled} & 41.478 & 0.0277 & 0.9940 \\
 F-\ourmodel{} & 45.618 & 0.0171 & 0.9987 \\
 L-\ourmodel{} & 43.837 & 0.0210 & 0.9982 \\
\midrule
\multicolumn{4}{c}{New Geometry}\\
\midrule
 Ne FBP-R & 34.984 & 0.0812 & 0.9829 \\
 Li FBP-R & 36.137 & 0.0711 & 0.9854 \\
 Li FBP-H & 36.243 & 0.0702 & 0.9833 \\
 DSigNet~\cite{he2021downsampled} & 29.012 & 0.0250 & 0.9409 \\
 F-\ourmodel{} & 36.236 & 0.0728 & 0.9863 \\
 L-\ourmodel{} & 36.658 & 0.0669 & 0.9864 \\
\bottomrule[1pt]
\end{tabular}\vspace{-2mm}
\end{center}
\end{table}}

\def\arraystretch{1.1}
\renewcommand{\tabcolsep}{3 pt}{
\begin{table}
\caption{Subjective quality score (Mean ± SD) of reconstructions for different methods on clinical LDCT dataset.} \vspace{-4mm}
\label{tab:score}
\small
\begin{center}
\begin{tabular}{lccc}
\toprule[1pt]
 Method & Noise Reduction & Structure Preservation & Overall Quality  \\
\midrule
Lowdose & 2.23$\pm$0.51 & 2.61$\pm$0.58 & 2.38$\pm$0.54 \\
 Ne FBP-R & 2.97$\pm$0.45 & 3.19$\pm$0.39 & 3.15$\pm$0.36 \\
 Li FBP-R & 3.85$\pm$0.31 & 3.77$\pm$0.29 & 3.79$\pm$0.32 \\
 L-\ourmodel{} & 4.19$\pm$0.26 & 4.19$\pm$0.26 & 4.19$\pm$0.26 \\
 F-\ourmodel{} & 4.26$\pm$0.30 & 4.25$\pm$0.26 & 4.24$\pm$0.22 \\
\bottomrule[1pt]
\end{tabular}\vspace{-2mm}
\end{center}
\end{table}}

\subsection{Generalization capability of the proposed \ourmodel{}}\label{sec:experiments}

We further illustrate the reconstruction performance and generalization capability of the \ourmodel{} in two scenarios: (i) cross-domain cases, where the imaging operators mismatch between the training and testing sets, and {(ii) domain-shift cases, where the imaging operators are the same but the image distributions differ, \textit{i.e.}, different CT datasets will be used for training and testing.} 

(i) Cross-domain cases. The imaging operators of the training and testing sets are listed in Table~\ref{tab:geometries}. Note that the proposed \ourmodel{} is unaware of the imaging operators of testing data in the training stage. Table~\ref{tab:different} shows the quantitative evaluations of the reconstruction performance of competing methods on testing data. \HUI{The proposed F-\ourmodel{} and L-\ourmodel{} consistently achieve relatively higher CT reconstruction accuracy than traditional FBP, demonstrating the fine generalization capability against shifts in the imaging operator.} In contrast, DSigNet~\cite{he2021downsampled}, which relies on complex black-box networks to learn the backprojection process, fails to deliver satisfactory results. It is worth noting that L-\ourmodel{}, with linear basis functions, outperforms the F-\ourmodel{} with Fourier basis functions. We attribute this to the fact that the linear bases are more reliant on coordinate positions and better at capturing patterns in neighboring coordinates. 

(ii) Domain-shift cases. {In this study, Clinical LDCT dataset is selected as the testing set, while the AAPM Mayo Clinic dataset still remains as the training set for our model. Although the imaging operators of the training and testing sets remain the same, the CT images originate from different scanners. For comparison, we applied the traditional FBP algorithm directly to the Clinical LDCT dataset to obtain the reconstruction results.} Fig.~\ref{fig:lesions} shows the representative reconstructions with lesions using different methods. It can be seen that the proposed F-\ourmodel{} and L-\ourmodel{} methods achieve higher fidelity reconstructions with clear tissue depiction, \textit{i.e.}, better preserving lesion details. Moreover, subjective assessments from experienced radiologists are also conduced to evaluate the potential clinical application. Table~\ref{tab:score} lists the average scores, indicating the potential application of the proposed \ourmodel{} in clinical settings. These results highlight the reliability of the proposed \ourmodel{} against variations in imaging operators and domain shifts.


\renewcommand{\tabcolsep}{2 pt}{
\begin{figure}[t!]
	\begin{center}
			\includegraphics[width=0.98\linewidth]{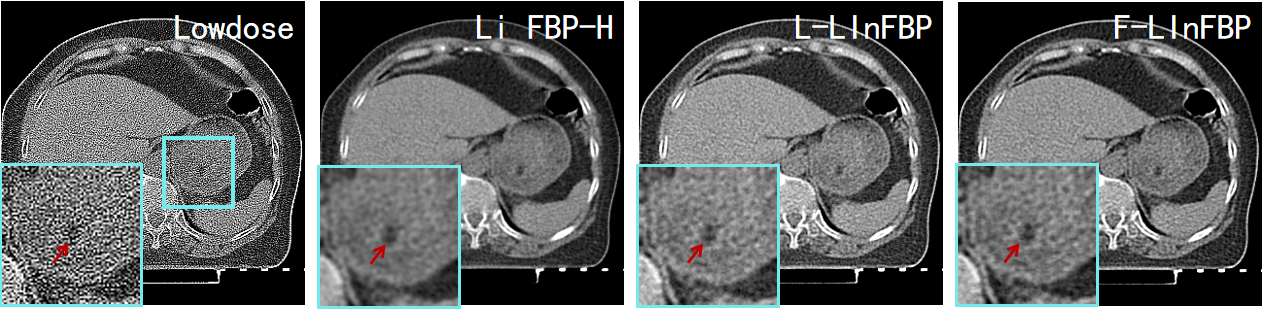}
   
		\caption{{Representative reconstructions with lesions using different methods in domain-shift experiments where different CT datasets are used for training and testing.}}
		\label{fig:lesions}
	\end{center}
 \vspace{-4mm}
\end{figure}}

\subsection{Running cost evaluation}\label{sec:costs}

Table~\ref{tab:costs} reports a comparison of floating point operations (FLOPs) and corresponding memory cost for reconstructions at 290 views. Notably, when compared to F-\ourmodel{}, L-\ourmodel{} with even more basis functions (5 compared to 3), the table demonstrates an obvious reduction in both FLOPs and memory cost. It's important to highlight that while \ourmodel{} introduces some additional computational costs compared to traditional FBP, such increase in FLOPs should be unsubstantial when compared to mainstream deep learning methods. This underscores the practicality and flexible plug-and-play capability of our method (especially L-\ourmodel{}).

\def\arraystretch{1.1}
\renewcommand{\tabcolsep}{6 pt}{
\begin{table}
\caption{Comparisons of GFLOPs and memory cost for different methods.}
\label{tab:costs}
\small
\begin{center}
\begin{tabular}{lcc}
\toprule[1pt]
 Method & GFLOPs (G) & Memory Costs (GB) \\
\midrule
 Nearest FBP & 167.6 & 2,424.2 \\
 Cubic FBP & 552.2 & 3,222.3 \\
 Linear FBP & 395.7 & 3,482.3 \\
 L-\ourmodel{} & 742.3 & 3,481.5 \\
 F-\ourmodel{} & 2,518.7 & 12,767.2 \\
\midrule
 Ori iRadonMap & 411,127.4 & 2,615.5 \\
 iRadonMap + L-\ourmodel{} & 411,702.1 & 3,485.5 \\
 iRadonMap + F-\ourmodel{} & 413,478.4 & 12,771.2 \\
 Ori DICDNet & 1,110,232.9 & 2,618.8 \\
 DICDNet + L-\ourmodel{} & 1,110,807.7 & 3,488.8 \\
 DICDNet + F-\ourmodel{} & 1,112,584.0 & 12,774.5 \\
\bottomrule[1pt]
\end{tabular}
\end{center}
\end{table}}

\subsection{Influence of the number of basis functions}\label{sec:influence}
In this section, we investigate the influence of the number of basis functions on the reconstruction performance and computational costs. The experimental results are shown in Fig.~\ref{fig:influence}. Due to the higher training costs associated with F-\ourmodel{}, we only investigate the influence of its first four basis functions. {It can be observed that the computational costs of F-\ourmodel{} significantly increases as the number of basis increases, far exceeding that of FBP.} However, L-\ourmodel{} can maintain a relatively lower training costs, with FLOPs increasing only slightly. Furthermore, F-\ourmodel{} exhibits degraded reconstruction performance after the number of basis functions increases, possibly due to the training complexity caused by a larger number of parameters. {In contrast, in the observed experiments with different basis numbers, L-\ourmodel{} consistently obtain improved reconstruction performance as the basis number increases.} Overall, both methods consistently outperform the FBP with clear margins, demonstrating the robustness of the proposed \ourmodel{}.

\renewcommand{\tabcolsep}{2 pt}{
\begin{figure}[t!]
	\begin{center}
		\begin{tabular}{cc}
			
			\includegraphics[height=0.42\linewidth]{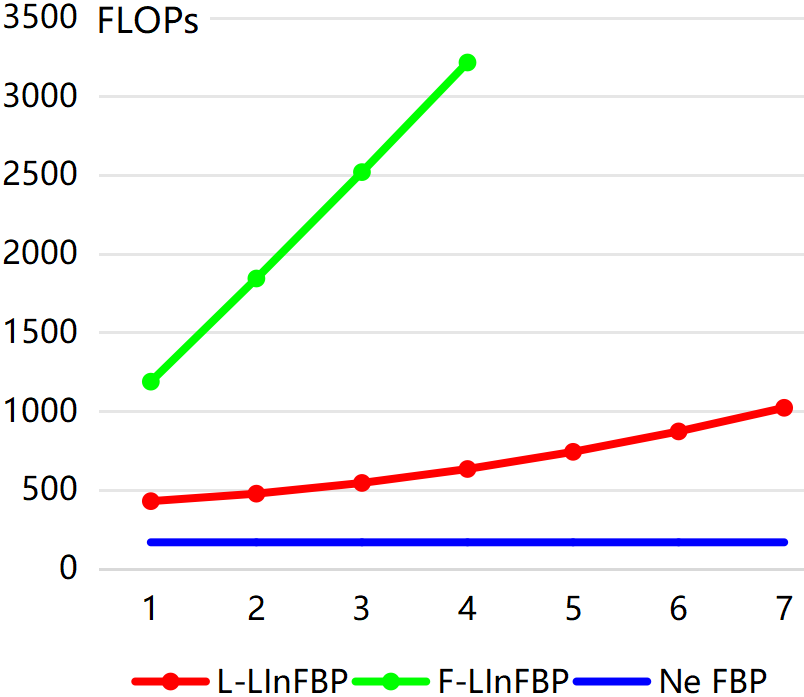}  &
			\includegraphics[height=0.42\linewidth]{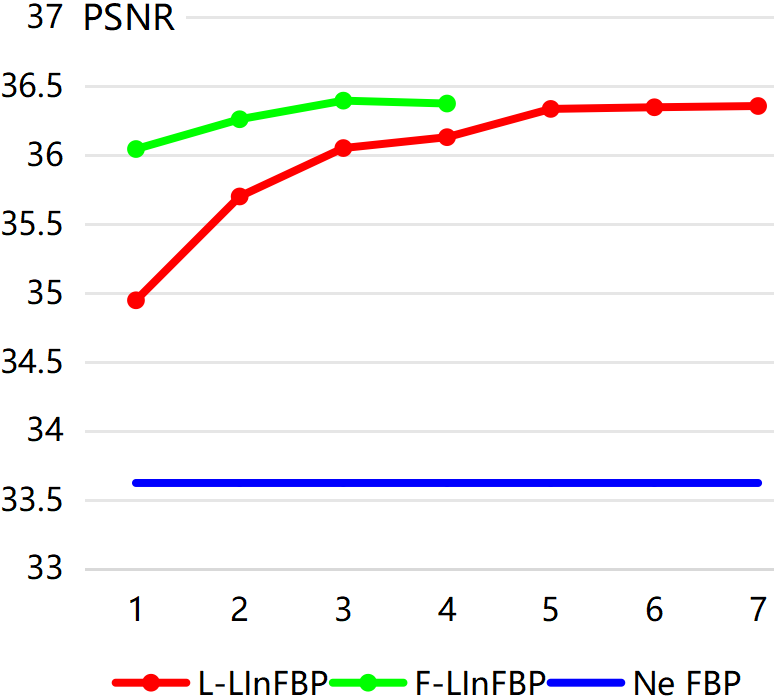} \\
   
		\end{tabular}
		\vspace{-2mm}
		\caption{Training costs (FLOPs) and reconstruction performance (PSNR) for different methods under different number of basis functions.}
		\label{fig:influence}
	\end{center} \vspace{-5mm}
\end{figure}}


\subsection{Influence of different loss functions}

{In this section, we investigate the impact of different loss functions on the CT reconstruction performance of our model. In addition to the commonly used L2 loss in Equation (\ref{eq:loss_I}), we incorporate another popular loss function, the image gradient difference loss (GDL), which is able to help reduce noise and preserve structural details~\cite{nie2018medical,mathieu2016deep}. The general form of the loss function can be written as}
\begin{equation}
	\mathcal{L}(I^*, I) = \Vert I^* - I \Vert^2_2 + \lambda(\Vert \mathbf{D_1} I^* - \mathbf{D_1} I \Vert_1 + \Vert \mathbf{D_2} I^* - \mathbf{D_2} I \Vert_1),
\end{equation}
{where $\mathbf{D_1}$ and $\mathbf{D_2}$ denote the finite difference operation along the first and second dimension of the 2D image $I$ with appropriate boundary conditions. We adjust the parameter $\lambda$ to study the influence of these two loss functions on the model. The experimental results under the quarter-view condition are presented in the Table~\ref{tab:loss}.}

\def\arraystretch{1.1}
\renewcommand{\tabcolsep}{12 pt}{
\begin{table}
\caption{{Quarter-view reconstruction performance of our methods under different loss combination weights.}}\label{tab:loss} 
\small
\begin{center}
\begin{tabular}{c|c|c|c}
\toprule[1pt]
 $\lambda$ & Method & PSNR & \HUI{FSIM}  \\
\midrule
\multirow{2}{*}{0} & F-\ourmodel{} & \textbf{36.397} & \HUI{\textbf{0.9702}} \\
 & L-\ourmodel{} & \textbf{36.337} & \HUI{0.9688} \\
 \midrule
 \multirow{2}{*}{$1.0 \times 10^{-3}$} & F-\ourmodel{} & 36.321 & \HUI{0.9701} \\
 & L-\ourmodel{} & 36.319 & \HUI{\textbf{0.9710}} \\
\midrule
 \multirow{2}{*}{$1.0 \times 10^{-2}$} & F-\ourmodel{} & 35.641 & \HUI{0.9699} \\
 & L-\ourmodel{} & 35.644 & \HUI{0.9707} \\
 \midrule
 \multirow{2}{*}{$1.0 \times 10^{-1}$} & F-\ourmodel{} & 34.865 & \HUI{0.9670} \\
 & L-\ourmodel{} & 34.811 & \HUI{0.9677} \\
 \midrule
 \multirow{2}{*}{$1.0 \times 10^{0}$} & F-\ourmodel{} & 32.596 & \HUI{0.9635} \\
 & L-\ourmodel{} & 32.723 & \HUI{0.9638} \\
 \midrule
  \multirow{2}{*}{$1.0 \times 10^{1}$} & F-\ourmodel{} & 30.986 & \HUI{0.9613} \\
 & L-\ourmodel{} & 30.889 & \HUI{0.9624} \\
 
\bottomrule[1pt]
\end{tabular}\vspace{-2mm}
\end{center}
\end{table}}

{The experimental results reveal a surprising finding: as the value of $\lambda$ increases, which indicates that more weight is places on the GDL, the model's accuracy decreases for both F-\ourmodel{} and L-\ourmodel{}. The best CT reconstruction accuracy is achieved when $\lambda$ is set to 0, meaning only the L2 loss is used. This suggests that, for deep learning models for CT reconstruction, L2 loss still remains one of the most effective loss functions compared to GDL. Therefore, in our experiments, we maintain the loss function in Equation (\ref{eq:loss_I}) as our final objective.}

\subsection{Influence of different detector bin width}

{To further evaluate the differences between our method and traditional FBP approaches under varying geometries, we conducted experiments to study the impact of different detector bin widths on reconstruction accuracy. Using the quarter-view and original geometries in Section~\ref{sec:aapm} as the baseline, we kept all other parameters fixed as constants while proportionally adjusting the number of detector bins and bin widths. The reconstruction accuracy for both traditional FBP methods and our proposed \ourmodel{} is shown in Table~\ref{tab:bin width}.}

\def\arraystretch{1.1}
\renewcommand{\tabcolsep}{7 pt}{
\begin{table}
\caption{{Influence of different detector bin width on CT reconstructions for different methods under the quarter-view condition.}}
\label{tab:bin width}
\small
\begin{center}
\begin{tabular}{lccc}
\toprule[1pt]
Method & PSNR & NMSE & FSIM \\
\midrule
\multicolumn{4}{c}{{1472 detectors with bin width 0.6848 mm}}\\
\midrule
{Ne FBP-R} & 32.458$\pm$0.90 & 0.0782$\pm$0.009 & 0.9420$\pm$0.009 \\
{Li FBP-R} & 34.164$\pm$0.93 & 0.0643$\pm$0.008 & 0.9552$\pm$0.007 \\
{F-\ourmodel{}} & 36.529$\pm$0.95 & 0.0490$\pm$0.006 & 0.9711$\pm$0.005 \\
{L-\ourmodel{}} & 36.430$\pm$0.96 & 0.0496$\pm$0.006 & 0.9699$\pm$0.006 \\
\midrule
\multicolumn{4}{c}{{1104 detectors with bin width 0.9131 mm}}\\
\midrule
{Ne FBP-R} & 33.398$\pm$0.92 & 0.0702$\pm$0.008 & 0.9568$\pm$0.007 \\
 {Li FBP-R} & 34.690$\pm$0.94 & 0.0605$\pm$0.007 & 0.9630$\pm$0.007 \\
 {F-\ourmodel{}} & 36.457$\pm$0.93 & 0.0494$\pm$0.006 & 0.9718$\pm$0.005 \\
 {L-\ourmodel{}} & 36.398$\pm$0.92 & 0.0497$\pm$0.006 & 0.9722$\pm$0.005 \\
\midrule
\multicolumn{4}{c}{{736 detectors with bin width 1.3696 mm}}\\
\midrule
{Ne FBP-R} & 33.624$\pm$0.92 & 0.0684$\pm$0.008 & 0.9417$\pm$0.007 \\
{Li FBP-R} & 35.004$\pm$0.93 & 0.0584$\pm$0.007 & 0.9463$\pm$0.006 \\
{F-\ourmodel{}} & 36.397$\pm$0.94 & 0.0498$\pm$0.006 & 0.9702$\pm$0.005 \\
{L-\ourmodel{}} & 36.337$\pm$0.94 & 0.0501$\pm$0.006 & 0.9688$\pm$0.005 \\
\midrule
\multicolumn{4}{c}{{552 detectors with bin width 1.8261 mm}}\\
\midrule
{Ne FBP-R} & 32.968$\pm$0.92 & 0.0738$\pm$0.009 & 0.9591$\pm$0.007 \\
{Li FBP-R} & 34.510$\pm$0.92 & 0.0618$\pm$0.008 & 0.9632$\pm$0.006 \\
{F-\ourmodel{}} & 36.279$\pm$0.93 & 0.0504$\pm$0.006 & 0.9715$\pm$0.005 \\
{L-\ourmodel{}} & 36.157$\pm$0.94 & 0.0511$\pm$0.006 & 0.9698$\pm$0.005 \\
\midrule
\multicolumn{4}{c}{{368 detectors with bin width 2.7392 mm}}\\
\midrule
{Ne FBP-R} & 31.347$\pm$0.92 & 0.0890$\pm$0.011 & 0.9493$\pm$0.008 \\
{Li FBP-R} & 32.681$\pm$0.91 & 0.0763$\pm$0.009 & 0.9533$\pm$0.007 \\
{F-\ourmodel{}} & 35.971$\pm$0.94 & 0.0522$\pm$0.007 & 0.9677$\pm$0.006 \\
{L-\ourmodel{}} & 35.654$\pm$0.94 & 0.0542$\pm$0.007 & 0.9673$\pm$0.006 \\
\bottomrule[1pt]
\end{tabular}
\end{center}
\end{table}}

{From the table, it can be seen that when the number of detectors decreases and the detector bin width increases, the reconstruction accuracy of FBP with nearest and linear interpolation naturally declines. Our method also shows a slight reduction in accuracy under these conditions. However, the decrease is relatively smaller compared to traditional FBP methods, and numerically, the performance gap between our method and traditional FBP widens. This indicates that the interpolation issue becomes increasingly pronounced in cases of sparse detectors, and our method effectively mitigates this problem, resulting in only an unsubstantial accuracy reduction compared to the complete 736 detector condition.}

{Conversely, as the detector density increases, \textit{i.e.}, the number of detectors increases and the bin width decreases, the reconstruction accuracy of traditional FBP methods still declines. We hypothesize that this occurs because the image size remains unchanged, making each image pixel being shared by multiple detectors. As a result, the pixel values are divided across different detectors during the forward projection, introducing errors. To mitigate this, more precise forward projection interpolation algorithms are needed, as simple methods like ray tracing and distance-driven approaches fail to provide the required accuracy. In contrast to traditional FBP, our model shows a slight improvement in accuracy. This indicates that the learnable structure of our model can effectively compensate for these errors, resulting in a more precise construction of the continuous detector representation.}

\subsection{Influence of different filters}

{In this section, we aim to evaluate the impact of different filter functions. We conducted experiments under quarter-view conditions to analyze the reconstruction variations of both our method and traditional FBP as different backprojection filters were applied. We select three filters for comparison: the Ramp filter, representing a sharp function; the Cosine filter, representing a medium function; and the Hann filter, representing a smooth function. The quantitative reconstruction results are presented in Table~\ref{tab:filter}.}

\def\arraystretch{1.1}
\renewcommand{\tabcolsep}{7 pt}{
\begin{table}
\caption{{Influence of different filters on CT reconstructions for different methods under the quarter-view condition.}}
\label{tab:filter}
\small
\begin{center}
\begin{tabular}{lccc}
\toprule[1pt]
Method & PSNR & NMSE & FSIM \\
\midrule
\multicolumn{4}{c}{{Ramp filter}}\\
\midrule
{Ne FBP} & 33.624$\pm$0.92 & 0.0684$\pm$0.008 & 0.9417$\pm$0.007 \\
{Li FBP} & 35.004$\pm$0.93 & 0.0584$\pm$0.007 & 0.9463$\pm$0.006 \\
{F-\ourmodel{}} & 36.397$\pm$0.94 & 0.0498$\pm$0.006 & 0.9702$\pm$0.005 \\
{L-\ourmodel{}} & 36.337$\pm$0.94 & 0.0501$\pm$0.006 & 0.9688$\pm$0.005 \\
\midrule
\multicolumn{4}{c}{{Cosine filter}}\\
\midrule
{Ne FBP} & 34.598$\pm$0.93 & 0.0612$\pm$0.007 & 0.9643$\pm$0.006 \\
 {Li FBP} & 35.394$\pm$0.93 & 0.0558$\pm$0.007 & 0.9667$\pm$0.006 \\
 {F-\ourmodel{}} & 36.075$\pm$0.91 & 0.0514$\pm$0.009 & 0.9700$\pm$0.006 \\
 {L-\ourmodel{}} & 36.046$\pm$0.90 & 0.0516$\pm$0.009 & 0.9703$\pm$0.006 \\
\midrule
\multicolumn{4}{c}{{Hann filter}}\\
\midrule
{Ne FBP} & 34.711$\pm$0.92 & 0.0604$\pm$0.007 & 0.9655$\pm$0.006 \\
{Li FBP} & 35.205$\pm$0.92 & 0.0570$\pm$0.007 & 0.9670$\pm$0.005 \\
{F-\ourmodel{}} & 36.072$\pm$0.90 & 0.0514$\pm$0.009 & 0.9698$\pm$0.006 \\
{L-\ourmodel{}} & 36.045$\pm$0.90 & 0.0516$\pm$0.009 & 0.9688$\pm$0.007 \\
\bottomrule[1pt]
\end{tabular}
\end{center}
\end{table}}

{From the experimental results, it can be observed that under the quarter-view condition, the reconstruction accuracy of traditional FBP methods including both nearest and linear interpolation, is clearly dependent on the filter applied. The Cosine and Hann filters, representing medium and smooth functions, generally achieve higher reconstruction accuracy than the Ramp filter. This is likely because sparse-view reconstructed images are more susceptible to artifacts and noise. Using smoother filter functions will mitigate these issues, leading to better results. For our method, the performance experiences a slight decline when using the Cosine and Hann filters. This decrease might be due to the fact that smoother filters incline to not only reduce noise but also sacrifice some fine details. Nonetheless, our method still consistently outperforms traditional FBP methods across all filter types. This demonstrates the robustness and effectiveness of our approach, as it can dynamically adjust its learnable interpolation function to adapt to varying filters, ensuring stable performance across different conditions.}

\subsection{Comparison with learnable backprojection method}

{In this section, we compare our methods with learnable backprojection matrix (named LBM module), which has been introduced in iRadonMap~\cite{he2020radon}. In this approach, after the backprojection step, the resulting tensor will be multiplied by a learnable backprojection matrix of size $M \times (512 * 512)$, where $M$ is the number of sampling views. The learnable matrix makes the backprojection process more smoother. The experiments are conducted under quarter view and normal dose conditions, and the quantitative results are shown in Table~\ref{tab:learnable matrix}.}

\def\arraystretch{1.1}
\renewcommand{\tabcolsep}{5 pt}{
\begin{table}
\caption{{Quantitative comparison of reconstructions with FBP methods with learnable backprojection matrix under quarter view and normal dose conditions.}}
\label{tab:learnable matrix}
\small
\begin{center}
\begin{tabular}{lccc}
\toprule[1pt]
Method & PSNR & NMSE & FSIM \\
\midrule
{Ne FBP} & 33.624$\pm$0.92 & 0.0684$\pm$0.008 & 0.9417$\pm$0.007 \\
{Li FBP} & 35.004$\pm$0.93 & 0.0584$\pm$0.007 & 0.9463$\pm$0.006 \\
{Ne FBP + LBM} & 33.635$\pm$0.93 & 0.0683$\pm$0.007 & 0.9606$\pm$0.006 \\
{Li FBP + LBM} & 35.011$\pm$0.93 & 0.0583$\pm$0.007 & 0.9647$\pm$0.005 \\
{F-\ourmodel{}} & 36.397$\pm$0.94 & 0.0498$\pm$0.006 & 0.9702$\pm$0.005 \\
{L-\ourmodel{}} & 36.337$\pm$0.94 & 0.0501$\pm$0.006 & 0.9688$\pm$0.005 \\
\bottomrule[1pt]
\end{tabular}
\end{center}
\end{table}}

{From the experimental results, it can be observed that for the traditional parameter-free FBP methods, including nearest and linear interpolation, incorporating the learnable backprojection matrix results in a slight improvement in CT reconstruction performance. Notably, the FSIM metric shows a clear increase, likely because the learnable matrix enhances the preservation of phase and gradient information in the final reconstruction. Overall, our \ourmodel{} method, both F-\ourmodel{} and L-\ourmodel{}, outperform FBP methods with LBM, further validating the effectiveness of our approach compared to such layer-based backprojection operator.}

\section{Discussion and Conclusion}\label{sec:conclusion}

In this paper, we have proposed \ourmodel{}, a novel approach that leverages continuous backprojection representation modeling to enhance the accuracy of CT image reconstruction. Particularly, \ourmodel{}  alleviates the interpolation error
  in a learnable manner,  without altering its efficient reconstruction framework and theoretic principles. Building upon these foundations, \ourmodel{} tend to maintain the precise part of the FBP process as derived in mathematical theories, while 
  reduce the interpolation errors and achieve more accurate CT image reconstruction
  by employing deep learning strategies to adaptively construct more suitable continuous representations.
Extensive experimental results demonstrate that \ourmodel{} outperforms classical reconstruction methods in terms of both visual inspection and quantitative assessment.  

{The superiority of our proposed model can be summarized in two key aspects. First, it is a learnable, parameterized framework that enables to learn a continuous function which will more accurately reflect the true sinogram from the training dataset, thereby reducing interpolation errors compared with manually selected interpolations. Second, our model is locally adaptive, generating unique continuous representations between each pair of detector bins. Unlike traditional interpolations, these representations are highly flexible, allowing them to effectively capture the variations in the sinogram where traditional functions may fail to provide accurate results.} 

Yet the presented work still has potential limitations. Although the L-\ourmodel{} is capable of evidently enhancing computational efficiency compared with F-\ourmodel{}, the memory costs still approximately 1.5 times than that of the most commonly used
conventional FBP method (Nearest FBP), as depicted in Tab.~\ref{tab:costs}. 
Exploring further computational cost reduction strategy should still one of our future researches.
In addition, although our analysis indicates that the errors in traditional FBP primarily arise from limited detectors and rotation angles, we did not specifically design the network to address sparse-view issues in this study. We will further explore characterizing continuous representations under sparse views to enhance the accuracy of the summation process in FBP. 

{Besides, it is worth noting that our model is both general and flexible, allowing both the kernel basis and the generation model of the latent encoded vector to be freely replaced. Both components play critical roles in improving performance. On the one hand, the choice of the basis function set affects the computational complexity of the framework and can also influence the representation capability of LCR. Specifically, for our proposed linear basis function set (\ref{Linear}), a set of uniformly distributed anchor points $\{-k,...,0,...,k\}$ is placed around each detector. The LCR then performs equivalently as linear operations between each pair of anchor points, which is essentially an extension of traditional linear interpolation~\cite{blu2004linear}. Consequently, when using different basis functions, such as a combination of triangular or quadratic functions, the continuous representation curve will fundamentally change and may be better suited for different tasks. The choice of optimal basis set remains an important open problem in the proposed framework, and we would explore it further in future research. On the other hand, the latent encoder plays a crucial role in learning the combination coefficients for LCR, ensuring that the framework is able to effectively capture the variations across different sinograms and construct high-quality CT images. In our current implementation, we employed a simple and straightforward deep learning model, $\operatorname{Net}_w$, to predict combination weights for continuous representations. \HUI{Notably, our model contains only around 300 parameters, which is negligible compared to the millions typically used in conventional deep neural networks. While this lightweight design ensures high efficiency and interpretability, we acknowledge that employing a more sophisticated network architecture could potentially further improve the accuracy in characterizing continuous representations. In future research, we will also delve more into the design of $\operatorname{Net}_w$ and explore more possibilities.}}

Despite the limitations mentioned above, we believe \ourmodel{} still offers a promising direction for accurate CT imaging in medical applications. Our method starts from the theoretical foundations of the FBP process, clearly analyzing approximations and errors in its three stages presented in discrete practice. On this basis, the future research can not only expand upon this study to further refine the characterization of continuous representations of the sinogram to address interpolation errors caused by limited detectors, but also explore issues in the filtering and summation processes from two other perspectives. {These approaches would facilitate a complementary relationship between deep learning and CT reconstruction, effectively combining the strengths of deep learning with the theoretic principles of FBP-based CT reconstruction. Notably, our model can also be extended to iterative algorithms, as the backprojection operator $A^\mathsf{T}$ often plays a crucial role in most iterative methods. The proposed learnable-interpolation-based backprojection can seamlessly replace the traditional backprojection operator, enabling its integration into iterative algorithms. Therefore, the insights and solutions proposed in this study are broadly applicable and demonstrate substantial versatility.}

{
\bibliographystyle{IEEEtran}
\bibliography{tmi}
}

\end{document}